\pdfoutput=1
\documentclass[a4paper]{article}

\usepackage{a4wide}
\RequirePackage{natbib}

\usepackage[pdftex,final,bookmarksnumbered,bookmarksopen=false,breaklinks,colorlinks]{hyperref}
\hypersetup{
	final,
	bookmarksnumbered=true,
	bookmarksopen=true,
	bookmarksopenlevel=0,
	unicode=false,
	pdftoolbar=true,
	pdfmenubar=true,
	pdffitwindow=false,
	pdftitle={},
	pdfdisplaydoctitle=true,
	pdftoolbar=true,
	pdfmenubar=true,
	pdflang={English},
	pdfauthor={},
	pdfsubject={arXiv paper},
	pdfcreator={},
	pdfproducer={},
	pdfkeywords={},
	pdfnewwindow=true,
	breaklinks=true,
	hidelinks,       
	linkcolor=black,
	citecolor=black,
	filecolor=magenta,
	urlcolor=cyan
}
\usepackage{url}
\usepackage[top=1.5cm,bottom=2cm,right=2.5cm,left=2.5cm]{geometry}
\usepackage{titling}

\usepackage{graphicx}%
\usepackage{adjustbox}
\usepackage{subcaption}
\usepackage{multirow}%
\usepackage{amsmath,amssymb,amsfonts}%
\usepackage{amsthm}%
\usepackage{mathrsfs}%
\usepackage[title]{appendix}%
\usepackage{xcolor}%
\usepackage{textcomp}%
\usepackage{manyfoot}%
\usepackage{booktabs}%
\usepackage{listings}%
\usepackage{pgffor}
\usepackage[percent]{overpic}
\usepackage{booktabs}   
\usepackage{multirow}   
\usepackage{makecell}
\usepackage{rotating}
\usepackage{caption}
\usepackage{enumitem}
\usepackage{comment}

\newcommand{\alphab}{\boldsymbol{\alpha}} 
\newcommand{\Ab}{\boldsymbol{A}} 
\newcommand{\xb}{\boldsymbol{x}} 
\newcommand{\Xb}{\boldsymbol{X}}
\newcommand{\yb}{\boldsymbol{y}}
\newcommand{\zb}{\boldsymbol{z}}
\newcommand{\eb}{\boldsymbol{e}}
\newcommand{\Yb}{\boldsymbol{Y}}
\newcommand{\Pb}{\boldsymbol{P}}
\newcommand{\Wb}{\boldsymbol{W}}
\newcommand{\gb}{\boldsymbol{g}}
\newcommand{\ub}{\boldsymbol{u}}
\newcommand{\vb}{\boldsymbol{v}}
\newcommand{\Jb}{\boldsymbol{J}}
\newcommand{\zetab}{\boldsymbol{\zeta}}
\newcommand{\pib}{\boldsymbol{\pi}} 
\newcommand{\mub}{\boldsymbol{\mu}}
\newcommand{\thetab}{\boldsymbol{\theta}} 
\newcommand{\Sigmab}{\boldsymbol{\Sigma}}
\newcommand{\etab}{\boldsymbol{\eta}}

\newcommand{\Dsf}{\mathsf{D}}
\newcommand{\Hess}{\mathsf{H}}
\newcommand{\gd}{\mathbf{g}}
\newcommand{\Ib}{\mathbf{I}}
\newcommand{\oneb}{\mathbf{1}}
\newcommand{\Rb}{\mathbb{R}} 
\newcommand{\rd}{\mathrm{d}} 
\newcommand{\Dir}{\mathrm{Dir}}
\newcommand{\Norm}{\mathcal{N}}

\newcommand{\alphabt}{{\widetilde{\alphab}}}
\newcommand{\tDsf}{\widetilde{\Dsf}}
\DeclareMathOperator{\tr}{tr}
\DeclareMathOperator{\vecop}{vec}
\DeclareMathOperator*{\argmax}{arg\,max}
\DeclareMathOperator*{\argmin}{arg\,min}

\newtheorem{lemma}{Lemma}
\newtheorem{theorem}{Theorem}
\newtheorem{proposition}[theorem]{Proposition}%
\newtheorem{corollary}[theorem]{Corollary}%

\newtheorem{algorithm}{Algorithm}

\allowdisplaybreaks

\title{Bandwidth rules for Dirichlet kernel density estimation\\ on the simplex}

\author{
  Jiajin Xie \thanks{Department of Statistics, University of Auckland, New Zealand.}
  \and
  Eduardo Garc\'ia-Portugu\'es \thanks{Department of Statistics, Universidad Carlos III de Madrid, Spain. 
  Corresponding author: \href{mailto:edgarcia@est-econ.uc3m.es}{edgarcia@est-econ.uc3m.es}.}
  \and
  Yong Wang \thanks{Department of Statistics, University of Auckland, New Zealand.
  }
}

\predate{}%
\postdate{}%
\date{}

\begin{document}

\maketitle

\begin{abstract}
Dirichlet kernels provide a natural tool for density estimation on the simplex, but data-driven bandwidth selection for them has been mostly limited to likelihood cross-validation (LCV). We develop bandwidth selectors for Dirichlet kernel estimators of densities and projected first derivatives for compositional data. Using a Dirichlet mixture as the reference model, we derive an exact representation of the mean integrated squared error (MISE) and define the exact MISE (EMI) selector. We also derive a closed-form least-squares cross-validation (LSCV) criterion and introduce the reference-smoothed least-squares (RSLS) selector, which replaces the unstable leave-one-out term of LSCV with the fitted Dirichlet mixture. For derivative estimation, we obtain projected derivatives of the Dirichlet density and their integrability conditions, extend the new bandwidth selectors, and carry out kernel mean shift clustering on the simplex. Simulations show that EMI is the most competitive selector for larger samples and Dirichlet-like targets, RSLS generally performs best for small samples and is more stable than LSCV for derivative estimation, and LCV remains competitive for non-Dirichlet targets in density estimation. Applications to wholesale customer expenditure and sectoral employment data illustrate density estimation and clustering, respectively.
\end{abstract}


\section{Introduction}
\label{sec:intro}

In this paper, we study bandwidth selection for kernel density estimation (KDE) with compositional data. The choice of bandwidth is critical for the resulting estimator, as it governs the trade-off between bias and variance and hence the overall smoothness of the estimate. Comprehensive references on kernel smoothing and bandwidth selection can be found in the monographs by \cite{Wand1995}, \cite{Scott2015}, and \cite{Chacon2018}. Density derivative estimation, and its associated bandwidth selection, has also attracted interest \citep{Horova2013, Chacon2018}, particularly due to its application to clustering \citep{Chacon2015}.

Compositional data, which consist of proportions of components within a whole, arise in many applied fields. Examples include sorting minerals by chemical composition in geosciences \citep{Buccianti2006}, studying how specific foods affect human health in nutrition research \citep{Leite2016}, and analyzing microbiome compositions and their effects \citep{Tsilimigras2016,Gloor2017,Calle2023}. The sample space of compositional data is the simplex, that is, the set of positive vectors with unit sum. Consequently, standard multivariate techniques are not appropriate, and specialized methods that account for their features are required. The classical reference for the statistical analysis of compositional data is \cite{Aitchison1986}. 

KDE for compositional data was first introduced by \cite{Aitchison1985}, who proposed two kernels, the Dirichlet and the additive log-ratio (alr) Gaussian kernel, along with the likelihood cross-validation (LCV) bandwidth selector. In particular, the Gaussian kernel is used in conjunction with the alr transformation, which maps compositions to the Euclidean space. \cite{Martin2006} proposed a centered variant of the Dirichlet kernel based on the Aitchison geometric transformation, and \cite{Chacon2011} subsequently introduced a scaled version of this kernel. \cite{Ouimet2022} incorporated Aitchison's Dirichlet kernel into a new framework. For its theoretical properties and related developments, see \cite{Bertin2023}, \cite{Daayeb2025}, and \cite{Daayeb2026}. For the log-ratio Gaussian kernel, \cite{Chacon2011} replaced alr with the isometric log-ratio (ilr) transformation, and obtained a more flexible ilr Gaussian kernel with a full-bandwidth matrix chosen by cross-validation \citep{Duong2005} or plug-in \citep{Chacon2010} methods. A different transformation was used by \cite{Yoon2025}, who mapped the compositions onto the hypersphere with a square-root map to handle zero components.

The aim of this work is to develop new bandwidth selectors for density and density derivative estimation of compositional data. We focus on the Dirichlet distribution as a kernel because it is naturally adapted to the simplex without requiring a log-ratio transformation. First, a bandwidth selector is proposed that uses an estimated Dirichlet mixture as the reference density. In this framework, a representation of the exact mean integrated squared error (MISE) is derived for Dirichlet mixtures, which serves as an objective function for bandwidth selection. A modified expectation-maximization (EM) algorithm is proposed for fitting Dirichlet mixtures, with the number of components selected by the Bayesian information criterion (BIC). This mixture-based selector is motivated by the multivariate approach of \cite{Cwik1997}. Using the same convolution identities, we further obtain a simple loss function for the least-squares cross-validation (LSCV) selector. To mitigate its numerical instability, we propose a new reference-smoothed least-squares (RSLS) selector by modifying the LSCV criterion. In addition, we study density derivative estimation on the simplex by introducing projected derivatives of the Dirichlet density, along with the corresponding integrability conditions. Finally, the proposed bandwidth selectors are extended to projected first-derivative estimation, with the corresponding exact MISE, LSCV and RSLS formulae derived, and this leads to a kernel mean shift clustering method.

The remainder of the paper is organized as follows. Section~\ref{sec:backg} reviews background on Dirichlet kernels and common bandwidth selectors. Section~\ref{sec:dir_mix} presents the Dirichlet mixture reference model and its estimation, while Section~\ref{sec:bs_density} introduces bandwidth selectors for Dirichlet KDE based on the exact MISE, LSCV, and RSLS criteria. Section~\ref{sec:derivative} develops projected derivatives of the Dirichlet density, establishes their integrability conditions, and introduces a kernel mean shift clustering method. Section~\ref{sec:bs_derivative} extends the proposed bandwidth selectors to projected first-derivative estimation. Section~\ref{sec:simu} contains simulation studies comparing Dirichlet kernel selectors for density and first-derivative estimation. Finally, Section~\ref{sec:app} illustrates real data applications to density estimation and clustering, and some conclusions are given in Section~\ref{sec:conclusion}. The technical lemmas, proofs, and additional simulation results are provided in the Appendix. The code replicating all the empirical results of the article is available at the GitHub repository \url{https://github.com/JiajinX/Dirichlet_KDE}.

\section{Background}
\label{sec:backg}

\subsection{Dirichlet KDE on the simplex}
\label{sec:Dir_KDE}

The sample space of compositional data $\xb = (x_1, \dots, x_D)^\top$ with $D:=d+1$ parts is the \emph{$d$-dimensional simplex}
\begin{align*}
    \Delta^d := \bigg\{\xb \in \mathbb{R}^D: x_1,\dots,x_D > 0, \, \sum_{k=1}^D x_k = 1 \bigg\}.
\end{align*}
We equip $\Delta^d$ with $\lambda_d(\rd\xb) := \rd x_1 \cdots \rd x_d$, the Lebesgue measure in the coordinates $(x_1, \ldots, x_d)$, with $x_D = 1 - \sum_{k=1}^d x_k$. Equivalently, let
\begin{align*}
    \Omega := \bigg\{\zb \in \Rb^d : z_1, \ldots, z_d > 0, \,\sum_{k=1}^d z_k < 1\bigg\}, ~~~~
    \iota(\zb) := \bigg(z_1, \ldots, z_d, 1 - \sum_{k=1}^d z_k \bigg)^\top,
\end{align*}
so that the affine map $\iota$ is a bijection from $\Omega$ onto $\Delta^d$. Then, for a function $\varphi$ on $\Delta^d$,
\begin{align}
    \int_{\Delta^d} \varphi(\xb)\,\lambda_d(\rd\xb) = \int_\Omega \varphi(\iota(\zb))\,\rd\zb.
    \label{eq:measure}
\end{align}

For compositional data, the widely used Dirichlet distribution, denoted by $\Dir(\alphab)$, has a density of the form
\begin{align}
    f_{\Dir}(\xb; \alphab) := 
    C_\Dir(\alphab)\prod_{k=1}^D x_k^{\alpha_k - 1},~~~~
    C_\Dir(\alphab) = \frac{\Gamma\Bigl(\sum_{k=1}^D \alpha_k\Bigr)}{\prod_{k = 1}^D\Gamma(\alpha_k)}, ~~~~
    \xb \in \Delta^d, ~\alphab \in \Rb^D_+. 
    \label{eq:dir}
\end{align}
The vector $\alphab = (\alpha_1, \dots, \alpha_D)^\top$ controls the location and shape of the distribution, with the concentration given by $\alpha_0 = \alphab^\top\mathbf{1}$. The distribution is log-concave if $\alphab \geq \oneb$ (element-wise), and has a unique mode at $\thetab = (\alphab - \oneb)/(\alpha_0-D)$ if at least one $\alpha_k>1$. For $\alphab = \oneb$, the distribution reduces to the uniform on $\Delta^d$, with a constant density $\Gamma(D)$. The Dirichlet distribution is often regarded as a canonical reference model on the simplex, much as the Gaussian distribution is for Euclidean data. Unlike the Gaussian distribution, the shape of a Dirichlet distribution is affected by its location. Even with $\alpha_0$ fixed, changing the individual components $\alpha_k$ alters the coordinate-wise variances. Therefore, for a given $\alpha_0$, a Dirichlet distribution centered near the boundary appears more concentrated than one centered in the interior.

An important advantage of the Dirichlet KDE is the aggregation property of the Dirichlet distribution and its extension to mixtures. Let $\Xb \sim \sum_{j=1}^m\pi_j\Dir(\alphab_j)$ as in \eqref{eq:fm} and let $\{S_1,\dots, S_k\}$ be a partition of $\{1,\dots,D\}$. The amalgamated composition also follows a Dirichlet mixture:
\begin{align}
    \biggl(\sum_{\ell \in S_1} X_\ell, \dots, \sum_{\ell \in S_k} X_\ell\biggr)
    \sim
    \sum_{j=1}^m \pi_j\Dir\biggl(\sum_{\ell \in S_1} \alpha_{j,\ell}, \dots, \sum_{\ell \in S_k} \alpha_{j,\ell} \biggr).
    \label{eq:amalgamation}
\end{align}
This property allows high-dimensional mixtures of Dirichlet distributions to be amalgamated onto a lower-dimensional simplex, such as $\Delta^2$ or $\Delta^3$, thereby facilitating visualization, as illustrated in Section~\ref{sec:app_wholesale}.

Given a compositional sample $\Xb_1, \dots, \Xb_n$, the basic form for the Dirichlet kernel density estimator at a point $\xb \in \Delta^d$ is
\begin{align}
    \hat{f}_h(\xb) := \frac{1}{n} \sum_{i=1}^n K(\xb, \Xb_i; h),
    \label{eq:kde_basicform}
\end{align}
where $K(\cdot, \cdot; h): \Delta^d \times \Delta^d \rightarrow \Rb$ denotes a kernel with bandwidth $h > 0$. As usual in kernel smoothing, bandwidth selection significantly affects the final estimate: a large $h$ yields an oversmoothed density estimate, whereas a small $h$ yields an undersmoothed estimate concentrated around the sample points.

Different forms of Dirichlet kernels have been considered in \eqref{eq:kde_basicform}. The classical version was introduced by \cite{Aitchison1985}, as given by
\begin{align}
    K_\mathrm{At}(\xb, \Xb; h) := f_\Dir\left(\xb; \frac{\Xb}{h} + \mathbf{1}\right).
    \label{eq:Dirkernel_At}
\end{align}
Thus each Dirichlet kernel has parameter $\alphab = \Xb/h + \mathbf{1}$. Since $\Xb\in\Delta^d$ and $h>0$, we have $\alphab > \oneb$. Therefore, the kernel is unimodal and log-concave. Since $\alpha_0 = \alphab^\top \mathbf{1} = 1/h + D$, the bandwidth $h$ plays the role of the concentration, while $\Xb$ is the mode of the kernel. To avoid the concentration being affected by the mode, \cite{Martin2006} proposed an alternative kernel as follows:
\begin{align*}
    K_\mathrm{Mt}(\xb, \Xb; h) = f_\Dir\left(\xb \oplus \Xb^{-1}; \frac{1}{hD}\mathbf{1}\right),
\end{align*}
where $\xb \oplus \Xb^{-1} := \mathcal{C}\,(x_1 X_1^{-1}, \dots, x_D X_D^{-1})$, and $\mathcal{C}(x_1, \ldots, x_D) := (x_1 / \sum_{k=1}^D x_k, \dots, x_D / \sum_{k=1}^D x_k)$ is the closure operator that normalizes a positive vector to unit sum \citep{Aitchison1986}. Owing to the centering operation $\xb \oplus \Xb^{-1}$, the kernel remains centered at $\Xb$, with its shape and concentration depending only on $h$. In addition, the bandwidth is restricted to $0 < h < 1/D$ to ensure a unimodal and log-concave kernel shape. Building on this, \cite{Chacon2011} further introduced a scaled formulation
\begin{align*}
    K_\mathrm{Cc}(\xb, \Xb; h) = \frac{1}{h^{D-1}} f_\Dir\left(\frac{1}{h}\otimes(\xb\oplus\Xb^{-1}); \mathbf{1}+\frac{1}{D}\mathbf{1}\right),
\end{align*}
where $b \otimes\xb := \mathcal{C}\,(x_1^b,\dots, x_D^b)$. Compared with the previous definitions, the bandwidth no longer controls the kernel concentration, but acts through a scaling transformation of the composition, analogous to its role in the Gaussian kernel.

The Dirichlet kernels mentioned above are all based on the classical framework. \cite{Ouimet2022} reverses the roles of the evaluation point $\xb$ and the observation $\Xb_i$ in $K_\mathrm{At}$, leading to the estimator
\begin{align}
    \hat{f}_h(\xb) := \frac{1}{n} \sum_{i=1}^n K_\mathrm{At}(\Xb_i, \xb; h).
    \label{eq:associated_kernel}
\end{align}
The shape of the kernel varies with the position $\xb$ on the simplex. This variable smoothing allows the kernel to adapt near the boundary and avoids the boundary bias of classical kernel estimators \citep[Sect.~3]{Ouimet2022}. However, the resulting kernel estimator is not a proper density, since it does not integrate exactly to one.

The classical measure of error in KDE is the $L^2$ distance between the estimator $\hat{f}_h$ and the target density $f$, namely the integrated squared error (ISE). Since this is a random quantity that depends on the sample, one usually considers its expectation, the mean ISE (MISE):
\begin{align*}
    \mathrm{MISE}(h) = \mathbb{E}\bigl[\mathrm{ISE}[\hat{f}_h]\bigr] = \mathbb{E}\left[\int_{\Delta^d}\left(\hat{f}_h(\xb) - f(\xb)\right)^2 \,\lambda_d(\rd\xb)\right],
\end{align*}
which depends on the bandwidth $h$, the kernel $K$, the sample size $n$, and the target density $f$. A natural approach is therefore to select the bandwidth that minimizes the MISE:
\begin{align*}
    h_\mathrm{MISE} := \argmin_{h>0} \mathrm{MISE}(h).
\end{align*}
The MISE is often difficult to evaluate and optimize, as it typically lacks a tractable exact representation. In most cases, the asymptotic MISE (AMISE), which captures the dominating part of the MISE, is easier to analyze and optimize, while also providing an intuitive bias--variance trade-off.

\subsection{Bandwidth selectors}

The plug-in approach is a classical method for bandwidth selection. It minimizes the AMISE expression and plugging in estimates for the unknown curvature present in that formula. A particularly simple example is the rule-of-thumb method, which estimates the curvature by that of an estimated Gaussian \citep[p.~45]{Silverman1986}. To overcome the rigidity of the Gaussian fit, \cite{Cwik1997} used Gaussian mixtures as the reference density in the multivariate setting.

Cross-validation selectors adopt a different strategy for bandwidth selection. In particular, the least-squares cross-validation (LSCV) and likelihood cross-validation (LCV) selectors are defined as the minimizers of cross-validated estimates of the squared-error loss and the Kullback--Leibler loss, respectively, and can be written as
\begin{align}
    \hat{h}_\mathrm{LSCV} &:= \argmin_{h>0} \mathrm{LSCV}(h),~~~~\mathrm{LSCV}(h) := \int_{\Delta^d} \hat{f}_h(\xb)^2 \,\lambda_d(\rd\xb) - \frac{2}{n} \sum_{i=1}^n \hat{f}_{h}^{-i}(\Xb_i); \label{eq:lscvh}\\
    \hat{h}_\mathrm{LCV} &:= \argmax_{h>0} \mathrm{LCV}(h),~~~~~\mathrm{LCV}(h) := \sum_{i=1}^n \log \hat{f}_h^{-i}(\Xb_i), \nonumber
\end{align}
where $\hat{f}_h^{-i}$ represents the kernel estimator computed without the $i$-th observation. The LSCV criterion is motivated by the data-dependent part of the $\mathrm{ISE}[\hat{f}_h]$, namely
\begin{align}
    \int_{\Delta^d} \hat{f}_h(\xb)^2 \,\lambda_d(\rd\xb) - 2\int_{\Delta^d} \hat{f}_h(\xb)f(\xb) \,\lambda_d(\rd\xb). \label{eq:ISE_data_dependent}
\end{align}
It replaces the cross-product term by the leave-one-out average since $f$ is unknown. Consequently, $\mathrm{LSCV}(h)$ is an unbiased estimator of the MISE up to an additive constant. For Dirichlet KDE, \cite{Aitchison1985} used the LCV bandwidth selector, whereas \cite{Daayeb2025} employed the LSCV selector based on estimator \eqref{eq:associated_kernel}.

In this article, we focus on the classical kernel estimation framework. The estimator in \eqref{eq:associated_kernel} follows a different theoretical framework, and has already generated research on theoretical properties \citep{Ouimet2022, Bertin2023} and subsequent refinements \citep{Daayeb2025, Daayeb2026}. Since its introduction by \cite{Aitchison1985} and subsequent discussion in \cite{Chacon2011}, Dirichlet kernel smoothing within the classical framework has received comparatively limited attention and remains open to further investigation. Regarding the choice of kernel, we adopt $K_\mathrm{At}$ in \eqref{eq:Dirkernel_At} because of its simple and more exploitable form.

\section{Dirichlet mixture reference model and estimation}
\label{sec:dir_mix}

Since the proposed selectors use a Dirichlet mixture as a reference density, this section introduces the reference model and describes its estimation. A Dirichlet mixture with $m$ components is defined as
\begin{align}
    f_m(\xb) := \sum_{j=1}^m \pi_j f_\Dir(\xb; \alphab_j), ~~~~\sum_{j=1}^m \pi_j = 1, ~\pi_j > 0,~\alphab_j \geq \oneb,
    \label{eq:fm}
\end{align}
where $\alphab_j$ and $\pi_j$ denote the parameter vector and mixing weight of the $j$th component, respectively. We restrict $\alphab_j \geq \oneb$ to ensure each component is log-concave. The tuning parameter of \eqref{eq:fm} is the number of components $m$. The notation $f_{\widehat{m}}$ denotes the fitted mixture with $\widehat{m}$ components, where both $\widehat{m}$ and the component parameters are estimated from the data. 

The EM algorithm is commonly used to fit mixture models. For Dirichlet mixtures, \cite{Pal2022} introduced a hard EM algorithm for compositional clustering, while \cite{Xie2025} proposed a soft EM algorithm with a common concentration $\alpha_0$ across all components. We extend the method of \cite{Xie2025} by allowing each component to have its own concentration $\alpha_{0,j} = \alphab_j^\top \oneb$, with $\alpha_{0,j} \leq \bar{\alpha}_0$. This upper bound prevents individual components from collapsing onto single observations, which would otherwise lead to an unbounded likelihood.

Under the constraints $\alphab_j \geq \oneb$ and $\alpha_{0,j} \leq \bar{\alpha}_0$, the parameters $(\alphab_j, \pi_j)$ of the $j$th component are updated at each iteration of the EM algorithm as
\begin{align*}
    p_{ij} := \frac{\pi_j f_{ij}}{\sum_{l=1}^m \pi_l f_{il}},~~~~
    \pi_j' := \frac{1}{n} \sum_{i=1}^n p_{ij},~~~~
    \alphab_j' := \argmax_{\alphab_j} Q_j(\alphab_j), 
\end{align*}
where $f_{ij} := f_{\Dir}(\xb_i; \alphab_j)$, and primes denote updated quantities. Let $w_j=\sum_{i=1}^n p_{ij}$. Then
\begin{align*}
    Q_j(\alphab_j) &= \sum_{i=1}^n p_{ij} \Bigl[\log\Gamma(\alpha_{0,j})- \sum_{k=1}^D \log\Gamma(\alpha_{jk}) + \sum_{k=1}^D (\alpha_{jk} - 1) \log x_{ik}\Bigr]\\
    &= -w_j \Bigl[\,\sum_{k=1}^D\log\Gamma(\alpha_{jk}) - \log\Gamma(\alpha_{0,j})\Bigr] + \sum_{k=1}^D (\alpha_{jk} - 1) \sum_{i=1}^n p_{ij} \log x_{ik}.
\end{align*}
Since the square-bracketed expression in the second equality is the strictly convex log-partition function of the Dirichlet exponential family and $w_j > 0$, whereas the second term is affine in $\alphab_j$, $Q_j$ is strictly concave. Therefore, $Q_j$ has a unique global maximizer over the compact convex constraint set for $\alphab_j$. However, the maximizer has no closed form due to the log-gamma term. We therefore approximate $Q_j$ by its second-order Taylor series expansion about the current iterate $\alphab_j$, yielding 
\begin{align*}
    Q_j(\alphabt_j) \approx\; Q_j(\alphab_j) + \gd_j^\top(\alphabt_j - \alphab_j)
    + \frac{1}{2}(\alphabt_j - \alphab_j)^\top 
     \Hess_j (\alphabt_j - \alphab_j), ~~~~ \alphabt_j \geq \mathbf{1}, ~\alphabt_j^\top \oneb \leq \bar{\alpha}_0,
\end{align*}
where $\gd_j$ and $\Hess_j$ are the gradient vector and Hessian matrix of $Q_j$. Maximizing this quadratic approximation with respect to $\widetilde{\alphab}_j$ is equivalent to solving
\begin{align*}
    \argmax_{\widetilde{\alphab}_j}\,(\gd_j - \Hess_j \alphab_j )^\top \widetilde{\alphab}_j + \frac{1}{2} \widetilde{\alphab}_j^\top \,\Hess_j\, \widetilde{\alphab}_j, ~~~~ \alphabt_j \geq \mathbf{1}, ~\widetilde{\alphab}_j^\top \oneb \leq \bar{\alpha}_0.
\end{align*}
The upper-bound constraint on the concentration is handled using an active-set strategy. Specifically, we first solve the quadratic program under $\alphabt_j \geq \mathbf{1}$, treating $\widetilde{\alphab}_j^\top\oneb \leq \bar{\alpha}_0$ as inactive. If the resulting solution satisfies the upper bound, it is accepted. Otherwise, the constraint is active for that component, and we solve the quadratic programming again under both constraints. The quadratic programs can be solved efficiently using the R package \texttt{lsei} \citep{R-lsei}. To ensure the monotonic increase of $Q_j$ at each iteration, we employ a backtracking line search for $\alphabt_j$ based on the Armijo rule. Repeating the quadratic approximation and backtracking yields a sequence of updates for $\alphabt_j$ that approaches the exact maximizer $\alphab_j'$.

Given the number of mixture components $m$, we initialize the Dirichlet mixture using $k$-means clustering. A Dirichlet density is fitted to the observations within each cluster, and the resulting component densities are combined into a mixture, with the initial mixing proportions given by the relative cluster sizes. The EM algorithm then iteratively updates $(\alphab_1,\dots,\alphab_m)$ and $\pib$ until the increase in the log-likelihood $\ell$ falls below a specified tolerance.

To select the number of mixture components $m$, we use $\mathrm{BIC} = -2\ell + k \log n$, where $k= mD + m - 1$ is the number of free parameters. The selected model is the fitted mixture whose number of components minimizes the BIC. Specifically, we start by fitting mixtures with $m = 1, \dots, M$, computing the corresponding BIC values and obtaining $\widehat{m}$, the number of components with the smallest BIC. The search is continued until $\widehat{m}$ remains the minimizer over $m = 1, \dots, \widehat{m} + N_m$. Following the suggestion of \cite{GarciaPortugues2013}, we set $M = \lfloor \log n \rfloor$ and $N_m = 3$. The procedure is detailed as follows.

\begin{algorithm}[Mixture estimation with component selection]
\label{alg:fini_mix}
Let $\Xb_1,\dots,\Xb_n$ be a sample on $\Delta^d$.
\begin{enumerate}[label=\arabic*., leftmargin=3em, itemsep=0pt]
    \item Set $M=\lfloor\log n\rfloor$ and $\mathcal{M} = \{1, \dots, M\}$.
    \item \label{ag:step2} For $m \in \mathcal{M}$:
    \begin{enumerate}[label=\roman*., leftmargin=3em, itemsep=0pt]
        \item Estimate the $m$-component mixture using the EM algorithm.
        \item Compute the BIC of the fitted mixture.
    \end{enumerate}
    \item Let $\widehat{m}$ be the number of components of the
    mixture selected by the BIC.
    \item If $M < \widehat{m} + N_m$, set $\mathcal{M} = \{M+1, \dots, \widehat{m} + N_m\}$ and $M = \widehat{m} + N_m$ and return to Step~\ref{ag:step2} Otherwise, return $f_{\widehat{m}}$.
\end{enumerate}
\end{algorithm}

Note that, for density derivative estimation in Section~\ref{sec:derivative}, the stronger constraint $\alphab_j > c\oneb$ is required for the Dirichlet mixture components, where $c \geq r + \tfrac12$ is a constant and $r$ denotes the order of the derivative. This case can also be well handled by our algorithm.

\section{Bandwidth selection for density estimation}
\label{sec:bs_density}

\subsection{Exact mean integrated squared error for Dirichlet mixtures}

Inspired by \cite{Cwik1997}, who improved the rule-of-thumb method by using Gaussian mixtures as the reference density, and by \cite{GarciaPortugues2013}, who developed an analogous program for directional data using von Mises--Fisher mixtures, we propose a new bandwidth selector for compositional data using a Dirichlet mixture as the reference density. Instead of using AMISE for bandwidth selection, we consider MISE. As shown by \cite{Marron1992} in the univariate setting, the AMISE and MISE may differ substantially for moderate and even large sample sizes, leading to a significant discrepancy between $h_\mathrm{AMISE}$ and $h_\mathrm{MISE}$, which can become more pronounced as the dimension increases \citep{Chacon2026}. A substantial reduction in the estimation error may therefore be achieved by selecting the bandwidth to minimize the exact MISE rather than its asymptotic approximation. Fortunately, the exact MISE for \eqref{eq:Dirkernel_At} can be computed if the reference density is the Dirichlet mixture $f_m$, denoted by
\begin{align}
    \mathrm{MISE}_m(h) := \mathbb{E}\left[\int_{\Delta^d}\left(\hat{f}_h(\xb) - f_m(\xb)\right)^2 \,\lambda_d(\rd\xb)\right]. \label{eq:miseh}
\end{align}
The derivation is based on a convolution property of the Dirichlet distribution.

\begin{proposition} \label{prp:exact_mise}
Let $f_m$ be the density of an $m$-component Dirichlet mixture in \eqref{eq:fm}. The exact MISE of the density estimator based on the Dirichlet kernel \eqref{eq:Dirkernel_At}, for a random sample of size $n$ from $f_m$, is
\begin{align}
    \mathrm{MISE}_m(h) = n^{-1} C_d(h)+\boldsymbol{\pi}^\top\big[(1-n^{-1})\boldsymbol{\Psi}_2(h)-2\boldsymbol{\Psi}_1(h)+\boldsymbol{\Psi}_0(h)\big]\boldsymbol{\pi}, \label{eq:misedir}
\end{align}
where 
\begin{align*}
    C_d(h):=\sum_{j=1}^m \pi_j\int_{\Delta^d} \frac{C_\Dir(\yb/h+\oneb)^2}{C_\Dir(2\yb/h+\oneb)} f_\Dir(\yb;\alphab_j) \,\lambda_d(\rd\yb),~~\boldsymbol{\pi}=(\pi_1,\ldots,\pi_m)^\top.
\end{align*}
The $m\times m$ matrices $\boldsymbol{\Psi}_a(h)$ for $a=0, 1, 2$ have entries:
\begin{align*}
    \boldsymbol{\Psi}_0(h)&:=\left(\frac{C_\Dir(\alphab_i)C_\Dir(\alphab_j)}{C_\Dir(\alphab_i+\alphab_j-\oneb)}\right)_{ij},\\
    \boldsymbol{\Psi}_1(h)&:=\left(\int_{\Delta^d} \frac{C_\Dir(\yb/h+\oneb)C_\Dir(\alphab_i)}{C_\Dir(\yb/h+\alphab_i)} f_\Dir(\yb;\alphab_j)\,\lambda_d(\rd\yb)\right)_{ij},\\
    \boldsymbol{\Psi}_2(h)&:=\left(\int_{\Delta^d}\int_{\Delta^d}\frac{C_\Dir(\yb_1/h+\oneb)C_\Dir(\yb_2/h+\oneb)}{C_\Dir((\yb_1+\yb_2)/h+\oneb)}  f_\Dir(\yb_1;\alphab_i)f_\Dir(\yb_2;\alphab_j) \,\lambda_d(\rd\yb_1)\,\lambda_d(\rd\yb_2)\right)_{ij},
\end{align*}
where $C_\mathrm{Dir}$ is defined in \eqref{eq:dir}.
\end{proposition}

Note that all quantities in Proposition~\ref{prp:exact_mise} involving integrals can be written as expectations and approximated by Monte Carlo methods. More specifically, independent Monte Carlo samples are generated from each mixture component of $f_m$. These samples are held fixed and reused across all candidate $h$, yielding a smoother Monte Carlo objective and therefore reducing variability in bandwidth comparisons. Let $\Psi_{1,ij}(h)$ and $\Psi_{2,ij}(h)$ denote the $(i,j)$th entries of $\boldsymbol{\Psi}_1(h)$ and $\boldsymbol{\Psi}_2(h)$, respectively. The corresponding Monte Carlo approximations are
\begin{align*}
    C_d(h) &= \sum_{j=1}^m \pi_j \mathbb{E}_{\Yb^{(j)}\sim \Dir(\alphab_j)} \left[\frac{C_\Dir(\Yb^{(j)}/h + \oneb)^2}{C_\Dir(2\Yb^{(j)}/h + \oneb)}\right] \approx \sum_{j=1}^m \frac{\pi_j}{S} \sum_{s=1}^{S} \frac{C_\Dir(\Yb_s^{(j)}/h + \oneb)^2}{C_\Dir(2\Yb_s^{(j)}/h + \oneb)},\\
    \Psi_{1,ij}(h) &= \mathbb{E}_{\Yb^{(j)}\sim \Dir(\alphab_j)}\left[\frac{C_\Dir(\Yb^{(j)}/h+\oneb)C_\Dir(\alphab_i)}{C_\Dir(\Yb^{(j)}/h+\alphab_i)}\right]\approx \frac{1}{S}\sum_{s=1}^{S} \frac{C_\Dir(\Yb^{(j)}_s/h+\oneb)C_\Dir(\alphab_i)}{C_\Dir(\Yb^{(j)}_s/h+\alphab_i)},\\
    \Psi_{2,ij}(h) &= \mathbb{E}_{\left(\Yb^{(i,1)},\Yb^{(j,2)}\right)\sim\Dir(\alphab_i)\times\Dir(\alphab_j)}\left[\frac{C_\Dir(\Yb^{(i,1)}/h+\oneb)C_\Dir(\Yb^{(j,2)}/h+\oneb)}{C_\Dir((\Yb^{(i,1)}+\Yb^{(j,2)})/h+\oneb)} \right]\\
    &\approx \frac{1}{S}\sum_{s=1}^{S} \frac{C_\Dir(\Yb_{s}^{(i,1)}/h+\oneb)C_\Dir(\Yb_{s}^{(j,2)}/h+\oneb)}{C_\Dir((\Yb_{s}^{(i,1)}+\Yb_{s}^{(j,2)})/h+\oneb)}.
\end{align*}
Above $\Yb_s^{(j)}, s=1,\dots,S$ are independent and identically distributed (iid) draws from $\Dir(\alphab_j)$, where $S$ denotes the generated sample size for each component. For the approximation of $\Psi_{2,ij}(h)$, two independent Monte Carlo samples are required. In particular, when $i=j$, the two samples must still been drawn from the same distribution $\Dir(\alphab_i)$.

Monte Carlo evaluation of the integrals in \eqref{eq:misedir} offers a substantial computational advantage over direct Monte Carlo evaluation of \eqref{eq:miseh}. The generated samples under the former approach can be reused for different sample sizes $n$. By contrast, the latter approach requires generating $\xb_1^{(b)}, \dots, \xb_n^{(b)}$ for $b = 1, \ldots, B$, constructing the estimator for each dataset, and then approximating the integral by an \emph{additional} Monte Carlo step. This makes the former approach substantially more efficient, especially when $n$ is large. Section~\ref{subsec:exact_MISE} compares the two evaluations in terms of the MISE curves and computation times.

Building on the previous result, we introduce an exact MISE based bandwidth selector, denoted by EMI. The procedure is summarized below.

\begin{algorithm}[EMI selector]
    Let $\Xb_1,\dots,\Xb_n$ be a sample on $\Delta^d$ and consider the density estimator based on the Dirichlet kernel \eqref{eq:Dirkernel_At}.
    \begin{enumerate}[label=\arabic*., leftmargin=3em, itemsep=0pt]
        \item Compute a suitable estimate $\hat{f}_{\widehat{m}}$ using Algorithm~\ref{alg:fini_mix}.
        \item Obtain $\hat{h}_\mathrm{EMI} := \argmin_{h>0} \mathrm{MISE}_{\widehat{m}}(h)$.
    \end{enumerate}
    \label{alg:EMI}
\end{algorithm}

\subsection{Cross-validation}

Using similar convolution techniques, for the Dirichlet kernel, the LSCV selector admits a simpler expression for \eqref{eq:lscvh} that avoids explicit evaluation of the integral.

\begin{proposition} \label{prp:lscv}
    The LSCV of the density estimator based on the Dirichlet kernel \eqref{eq:Dirkernel_At} is
    \begin{align}
    \mathrm{LSCV}(h) &= \frac{1}{n^2}\sum_{i,j=1}^n \frac{C_{\Dir}(\Xb_i/h+\oneb)C_{\Dir}(\Xb_j/h+\oneb)}{C_{\Dir}((\Xb_i+\Xb_j)/h + \oneb)}-\frac{2}{n}\sum_{i=1}^n \hat{f}^{-i}_h(\Xb_i) \label{eq:lscv1}\\
    &= \frac{\Gamma(1/h+D)}{n^2}\sum_{i,j=1}^n \bigg\{
    \frac{\Gamma(1/h+D)}{\Gamma(2/h+D)} \prod_{k=1}^D \frac{\Gamma((X_{ik}+X_{jk})/h+1)}{\Gamma(X_{ik}/h+1)\Gamma(X_{jk}/h+1)}\nonumber\\
    &~~~~ -2(1-n^{-1})^{-1}(1-\delta_{ij})\prod_{k=1}^D \frac{X_{ik}^{X_{jk}/h}}{\Gamma(X_{jk}/h+1)}\bigg\}.\label{eq:lscv2}
\end{align}
\end{proposition}

Expression \eqref{eq:lscv1} provides a conceptually compact representation, whereas \eqref{eq:lscv2} gives an explicit form convenient for direct numerical evaluation. The overall evaluation cost is $O(n^2D)$, but the symmetry of the first double-sum term can be exploited to reduce its computational cost.

The leave-one-out average in LSCV can be sensitive to pairs of closely spaced observations, especially when the bandwidth $h$ is small. Such pairs can produce large values of $\hat f_h^{-i}(\Xb_i)$, causing sharp dips in $\mathrm{LSCV}(h)$. These dips can make bandwidth selection unstable and may lead to an overly small bandwidth \citep{Jones1996}. For the simplex, boundary effects can add to this problem, since Dirichlet kernels centered near the boundary are typically more peaked than those in the interior.

To alleviate this problem, we replace the unknown target density $f$ of \eqref{eq:ISE_data_dependent} by a smooth reference density $\tilde{f}$. This replacement introduces bias, so the resulting criterion is no longer an unbiased estimator of the data-dependent part of the MISE. However, it reduces variance caused by the leave-one-out average. When $\tilde{f}$ is close to $f$, this criterion should approximate the ideal criterion well. We choose a Dirichlet mixture density $f_m$ as $\tilde{f}$, so that the cross-product term can be computed in closed form. We refer to the resulting objective as the reference-smoothed least-squares (RSLS) criterion: 
\begin{align*}
    \mathrm{RSLS}_m(h) := \int_{\Delta^d} \hat{f}_h(\xb)^2 \,\lambda_d(\rd\xb) - 2 \int_{\Delta^d} \hat{f}_{h}(\xb)\,f_m(\xb) \,\lambda_d(\rd\xb).
\end{align*}

\begin{corollary} \label{coro:rsls}
    Let $f_m$ be the density of an $m$-component Dirichlet mixture in \eqref{eq:fm}. The RSLS of the density estimator based on the Dirichlet kernel \eqref{eq:Dirkernel_At} is
    \begin{align*}
        \mathrm{RSLS}_m(h) = \frac{1}{n^2}\sum_{i,j=1}^n \frac{C_{\Dir}(\Xb_i/h+\oneb)C_{\Dir}(\Xb_j/h+\oneb)}{C_{\Dir}((\Xb_i+\Xb_j)/h + \oneb)} - \frac{2}{n} \sum_{i=1}^n \sum_{j=1}^m \pi_j \frac{C_{\Dir}(\Xb_i/h+\oneb)C_{\Dir}(\alphab_j)}{C_{\Dir}(\Xb_i/h+\alphab_j)}.
    \end{align*}
\end{corollary}

The RSLS criterion is a hybrid between the LSCV and exact MISE selectors: it retains the data-driven nature of LSCV, while introducing a smooth reference density to reduce variability. The RSLS selector can be computed using Algorithm~\ref{alg:EMI}, with the exact MISE criterion replaced by the RSLS criterion.

\section{Density derivative estimation}
\label{sec:derivative}

\subsection{Projected derivatives and integrability}

Important structural features of the underlying target density are not fully captured by the density values alone but may be further revealed through the derivatives. For example, local minima and maxima are characterized by locations at which the first derivative is zero, and derivative estimation has several applications \citep[Chap.~6]{Chacon2018}. We first discuss density derivative estimation based on the Dirichlet kernel and then introduce bandwidth selection rules in Section~\ref{sec:bs_derivative}.

Let us state some required notation for derivatives of the Dirichlet density. Let $\Dsf = (\partial/\partial x_1, \dots, \partial/\partial x_D)^\top$ denote the differential operator with respect to $\xb$. The first derivative of the Dirichlet density, namely its gradient, is the $D$-dimensional (column) vector
\begin{align}
    \Dsf f_\Dir(\xb;\alphab) = \frac{\partial}{\partial \xb} f_\Dir(\xb; \alphab) =  \gb\,f_\Dir(\xb;\alphab),
    \label{eq:Dir_gradient}
\end{align}
where $\gb = \gb(\xb; \alphab):= (\alphab-\oneb)\oslash\xb$, and $\oslash$ denotes componentwise division. For higher-order derivatives, we use the vectorized representation by \citet[Sect.~2.6]{Chacon2018}. Specifically, let $\Dsf^{\otimes r} f_\Dir(\xb;\alphab) \in \Rb^{D^r}$ denote the $r$th derivative of $f_\Dir$ at $\xb$, defined as
\begin{align*}
    \Dsf^{\otimes r} f_\Dir(\xb;\alphab) := \frac{\partial^r f_\Dir(\xb;\alphab)}{(\partial \xb)^{\otimes r}},
\end{align*}
where $\otimes$ denotes the Kronecker product, and $\Ab^{\otimes r} = \Ab \otimes \stackrel{r}\cdots \otimes \Ab$ denotes the $r$th Kronecker power of a matrix $\Ab$. This object is a vector of length $D^r$ containing all partial derivatives of order $r$ of $f_\Dir$ at $\xb$. In particular, $\Dsf^{\otimes 0} f_\Dir = f_\Dir$ is the density itself, $\Dsf^{\otimes 1} f_\Dir = \Dsf f_\Dir$ is the gradient, and $\Dsf^{\otimes 2} f_\Dir = \vecop(\Hess f_\Dir)$, where $\Hess f_\Dir$ is the $D \times D$ Hessian of $f_\Dir$, with $\vecop(\cdot)$ denoting the column-stacking operator.

Using the definition of the first derivative in \eqref{eq:Dir_gradient} and Theorem 5(ii) of \cite{chacon-duong-2021}, we obtain a general expression for $\Dsf^{\otimes r} f_\Dir(\xb;\alphab)$. For $\xb \in \Delta^d$ and an integer $r\geq1$, the $r$th vectorized derivative of the Dirichlet density is
\begin{align}
    \mathsf{D}^{\otimes r} f_\Dir =
    (\mathbf{I}_D \otimes \mathcal{S}_{D,r-1}) \sum_{j=0}^{r-1} \binom{r-1}{j} \vecop\bigl\{\mathsf{D}^{\otimes (r-1-j)} f_\Dir \otimes \vecop_{D^j,D}^{-1} \mathsf{D}^{\otimes j} \gb\bigr\}.
    \label{eq:Dir_derivative}
\end{align}
Here, $\mathcal{S}_{D,r-1}$ is the $D^{r-1}\times D^{r-1}$ symmetrizer matrix, which symmetrizes Kronecker products and ensures uniqueness of the representation. An explicit form is given in \citet[p.~8]{chacon-duong-2021}. The term $\Dsf^{\otimes j}\gb$ appearing in \eqref{eq:Dir_derivative} is a vector of length $D^{j+1}$ and can be written as
\begin{align*}
    \mathsf{D}^{\otimes j} \gb = (-1)^j\,j!\,\sum_{k=1}^D \frac{\alpha_k - 1}{x_k^{j+1}} \eb_k^{\otimes (j+1)},
\end{align*}
where $\eb_k$ denotes the $k$th standard basis vector in $\Rb^D$. The operator $\vecop^{-1}_{D^j, D}$ transforms the above vector into a matrix of order $D^j \times D$. In particular, the derivatives for $r=2,3,4$ of the Dirichlet density are
\begin{align*}
    \Dsf^{\otimes 2} f_\Dir(\xb;\alphab) &= \left[\gb^{\otimes 2} + \Dsf \gb\right] f_\Dir(\xb; \alphab),\\
    \Dsf^{\otimes 3} f_\Dir(\xb;\alphab) &= \left[\gb^{\otimes 3} + 3\,\mathcal{S}_{D,3} (\gb \otimes \Dsf \gb) + \Dsf^{\otimes 2} \gb\right] f_\Dir(\xb; \alphab),\\
    \Dsf^{\otimes 4} f_\Dir(\xb;\alphab) &= \bigl[\gb^{\otimes 4} + 6\,\mathcal{S}_{D,4} \left(\gb^{\otimes 2}\otimes\Dsf\gb\right) + 3\,\mathcal{S}_{D,4} \left(\Dsf\gb\otimes\Dsf\gb\right) + 4\,\mathcal{S}_{D,4} \left(\gb\otimes\Dsf^{\otimes 2}\gb\right) + \Dsf^{\otimes 3}\gb \bigr] f_\Dir(\xb;\alphab).
\end{align*}

\begin{figure}[!tb]
    \centering
    \includegraphics[width=0.5\textwidth, trim=10cm 2.5cm 10cm 16cm, clip]{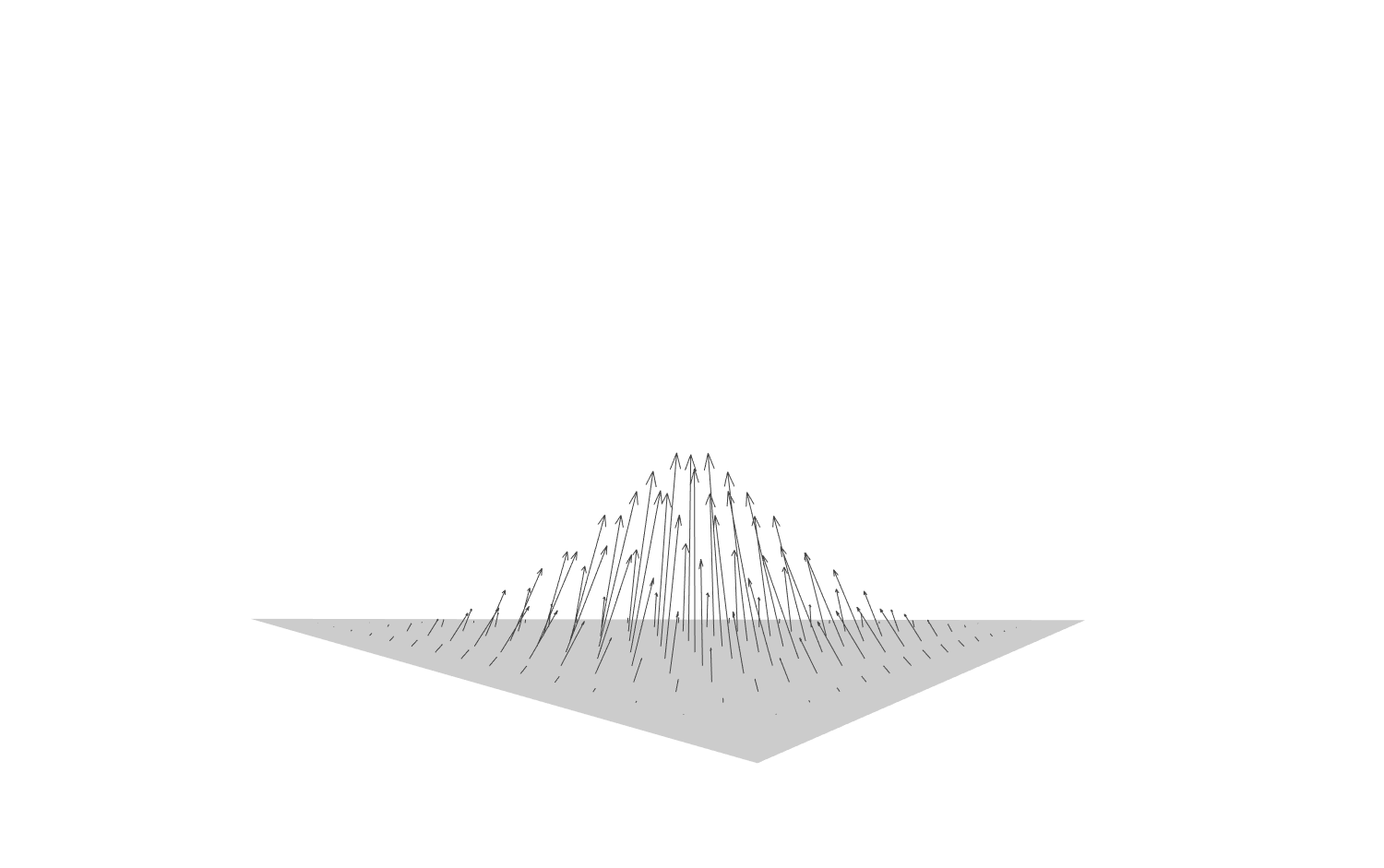}
    \hspace{0.05\textwidth}
    \includegraphics[width=0.4\textwidth, trim=1.3cm 1.4cm 1.2cm 2cm, clip]{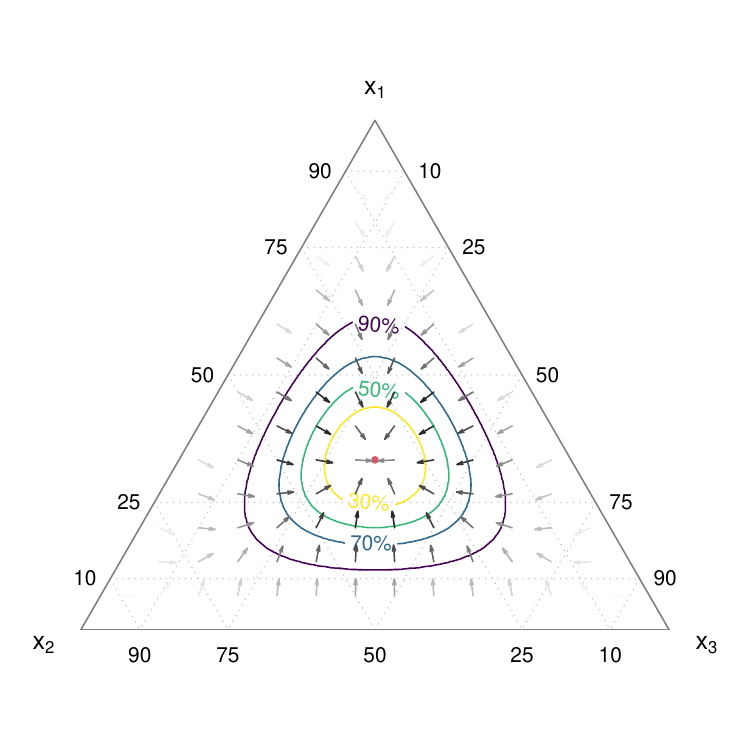}
    \caption{\small Unprojected gradient field of the $\Dir(5,5,5)$ density (left) and projected gradient field on $\Delta^2$ with density contours (right). The projected gradient vector fields are unit-normalized, with higher transparency indicating lower norm.}
    \label{fig:grad_field}
\end{figure}

Since the simplex $\Delta^d$ is embedded in $\Rb^D$, the ordinary derivatives of the Dirichlet density lie in $\Rb^D$. However, for analyzing the density and its structural features, only variations along directions tangent to the simplex are intrinsically meaningful. Therefore, we project the derivatives onto the tangent space $T_{\xb} \Delta^d = \{\mathbf{v} \in \Rb^D:\, \oneb^\top \mathbf{v} = 0\}$. Let $\Pb = \Ib_D - D^{-1} \oneb\oneb^\top$ be the $D\times D$ orthogonal projection matrix onto $T_{\xb} \Delta^d$, and define the projected differential operator $\tDsf := \Pb \Dsf$. Then, the $r$th projected derivative of $f_\Dir$ at $\xb$ is
\begin{align*}
    \tDsf^{\otimes r} f_\Dir(\xb;\alphab) = (\Pb\Dsf)^{\otimes r} f_\Dir(\xb;\alphab) = \Pb^{\otimes r}\Dsf^{\otimes r} f_\Dir(\xb;\alphab),
\end{align*}
which may be interpreted as the projection of the $r$th derivative onto $T_{\xb} \Delta^d$ through $\Pb^{\otimes r}$. Figure~\ref{fig:grad_field} illustrates the gradient field of the $\Dir(5,5,5)$ density and its projection onto $\Delta^2$.

The integrability of the Dirichlet derivatives depends on their order and parameter values. The proposition below provides sufficient conditions for the integrability of the $r$th derivative. In particular, $L^2$ integrability is required for the exact MISE, LSCV, and RSLS criteria to be finite.

\begin{proposition} \label{prp:integrability}
Let $r \geq 0$.
\begin{enumerate}[label=(\roman*)]
    \item \label{prop:integrability-i} If $\alpha_i>r$ for all $i=1,\dots,D$, then $\mathsf D^{\otimes r}f_\Dir\in L^1(\Delta^d)$.
    \item \label{prop:integrability-ii} If $\alpha_i>r+\tfrac12$ for all $i=1,\dots,D$, then $\mathsf D^{\otimes r}f_\Dir\in L^2(\Delta^d)$.
\end{enumerate}
\end{proposition}

We impose the sufficient condition $\alpha_k > r + \tfrac12$ on the kernel and mixture components. For density estimation ($r = 0$), the definitions in \eqref{eq:Dirkernel_At} and \eqref{eq:fm} already satisfy the condition because $\alphab \geq \oneb$. However, for density derivative estimation, $K_\mathrm{At}$ should be modified to
\begin{align}
    K_\mathrm{Atc}(\xb, \Xb; h) := f_\Dir\Bigl(\xb; \frac{\Xb}{h} + c\mathbf{1}\Bigr),
    \label{eq:Dirkernel_Atc}
\end{align}
where $c \geq r + \tfrac12$. Likewise, the condition $\alphab_j > (r + \tfrac12)\oneb$ is required for $f_m$. The same sufficient conditions apply to the projected derivative, since multiplication by $\Pb^{\otimes r}$ preserves $L^p$ integrability. Thus, the projected $r$th density derivative estimator is defined as 
\begin{align}
    \widehat{\tDsf^{\otimes r} f}_h(\xb) := \tDsf^{\otimes r}  \hat{f}_h(\xb) = \frac{1}{n}\sum_{i=1}^n \tDsf^{\otimes r} f_\Dir \Bigl(\xb; \frac{\Xb_i}{h} + c\oneb\Bigr).
    \label{eq:kde_derivative}
\end{align}

\subsection{Clustering} 
\label{sec:cluster}

An important application of density derivative estimation is clustering, where population clusters can be defined as the basins of attraction of the modes of the underlying density \citep{Chacon2015}. Let $\zetab_j$, $j=1,\dots,m$ denote the modes of a density $f$. A point $\xb$ is assigned to the $j$th cluster if the associated density-ascent trajectory converges to $\zetab_j$. The direction of maximum ascent at $\xb$ is the gradient $\Dsf f(\xb)$. Since the density-ascent trajectory of the compositions must remain on the simplex, we instead use the projected first derivative. Starting from $\xb_0 = \xb$, this leads to the iterative scheme
\begin{align*}
    \xb_{t+1} = \xb_t + \gamma\,\tDsf f(\xb_t), ~~~~ t = 0,\dots,N,
\end{align*}
where $\gamma > 0$ is the step size and $N$ is the maximum number of iterations. The iteration is terminated when $\tDsf f(\xb_t) \approx 0$, indicating that $\xb_t$ has approached a mode.

We make two modifications to this iteration scheme. First, we use the normalized projected gradient $\etab(\xb):=\tDsf f(\xb) / f(\xb)$ \citep{fukunaga-hostetler-1975}, so the trajectory moves more rapidly through low-density regions and more slowly through high-density regions. Second, we adapt the step size to ensure each update remains in the interior of the simplex. Although the projected gradient preserves the unit-sum constraint for $\xb_{t+1}$, a finite step may result in a negative component. Let $\eta_k(\xb_t)$ denote the $k$th component of $\etab(\xb_t)$. To ensure $x_{t,k} + a_t \, \eta_k(\xb_t) \geq 0$ for every $k$, the step size must satisfy
\begin{align*}
    a_t \leq a_{\max}(\xb_t) := \min_{k:\,\eta_k(\xb_t)<0} -\frac{x_{t,k}}{\eta_k(\xb_t)}.
\end{align*}
We therefore use the pointwise adaptive step size $a_t := \min\{a,\, \rho\,a_{\max}(\xb_t)\}$, where $\rho \in (0, 1)$ is a safety factor that prevents the updated point from reaching the boundary. 

Since the underlying density $f$ is unknown, we replace it by $\hat{f}_h$. This yields the iterative scheme
\begin{align}
    \xb_{t+1} = \xb_t + a_t \hat{\etab}_h(\xb_t), ~~~~ \hat{\etab}_h(\xb) := \frac{\tDsf \hat{f}_h(\xb)}{\hat{f}_h(\xb)}. \label{eq:cluster_iter}
\end{align}
Since the update depends directly on the gradient, a bandwidth selector tailored to first-derivative estimation is recommended \citep[Sect.~6.2.2]{Chacon2018}.

The number of population clusters is based on the modes of $\hat{f}_h$. Modal clustering can produce very small clusters due to outliers, local sampling variation, or a very small bandwidth. A post-processing step can merge these potentially spurious clusters with others based on the distance, density, or hierarchical linkage criteria. The complete clustering procedure is summarized below.

\begin{algorithm}[Kernel mean shift clustering on the simplex]
    Let $\Xb_1,\dots,\Xb_n$ be a sample on $\Delta^d$.
    \begin{enumerate}[label=\arabic*., leftmargin=3em, itemsep=0pt]
        \item Select a suitable bandwidth $h$ using a first-derivative bandwidth selector.
        \item For each $\Xb_i$, use the iterative scheme \eqref{eq:cluster_iter} until convergence to an endpoint $\yb_i$, $i=1,\dots,n$.
        \item Identify the distinct endpoints $\{\zetab_1,\dots,\zetab_m\}$ among $\{\yb_1,\dots,\yb_n\}$.
        \item Assign $\Xb_i$ to cluster $j$ if its trajectory converges to the mode $\zetab_j$.
        \item Merge spurious clusters using the post-processing step.
    \end{enumerate}
\end{algorithm}

\section{Bandwidth selection for first-derivative estimation}
\label{sec:bs_derivative}

With the definition of the projected $r$th density derivative estimator in \eqref{eq:kde_derivative}, we extend the proposed bandwidth selectors from density estimation to density derivative estimation.

\subsection{Exact mean integrated squared error for Dirichlet mixtures}
\label{subsec:exact_MISE}

The exact MISE corresponding to \eqref{eq:kde_derivative} is
\begin{align*}
    \mathrm{MISE}_{r,\,m}(h) :=   \mathbb{E}\left[\int_{\Delta^d}\Big\|\widehat{\tDsf^{\otimes r} f}_h(\xb) - \tDsf^{\otimes r} f_m(\xb)\Big\|^2 \,\lambda_d(\rd\xb)\right].
\end{align*}
This quantity admits an exact representation for the projected first derivative. For notational convenience, for $\boldsymbol{a},\boldsymbol{b}\in\mathbb{R}^D$, set
$\ub:=\boldsymbol a-\oneb$ and $\vb:=\boldsymbol b-\oneb$, and define
\begin{align}
    H(\boldsymbol{a}, \boldsymbol{b}) &:= (\oneb^\top\boldsymbol{a} + \oneb^\top\boldsymbol{b} - D - 1)^{\underline{2}} 
    \,\Big\{\frac{D-1}{D}\,\mathbf{1}^\top \left[\ub \odot \vb \oslash (\ub+\vb)^{\odot \underline{2}}\right] + \frac{1}{D} \,\mathbf{1}^\top\left[\ub \odot \vb \oslash (\ub+\vb)^{\odot 2}\right] \nonumber\\
    &~~~~ -\frac{1}{D}\mathbf{1}^\top\left[\ub \oslash (\ub+\vb)\right]\mathbf{1}^\top\left[\vb \oslash (\ub+\vb)\right]\Big\},
    \label{eq:Heq}
\end{align}
where $x^{\underline{2}} := x(x-1)$, $\xb^{\odot \underline{2}} := \xb \odot (\xb-\oneb)$, and $\odot$ denotes componentwise multiplication.

\begin{corollary} \label{coro:exact_mise_r1}
    Let $f_m$ be the $m$-component Dirichlet mixture in \eqref{eq:fm} with $\alphab_j>\tfrac32\oneb$, and let $c \geq \tfrac32$. The exact MISE of the projected first-derivative estimator based on the Dirichlet kernel \eqref{eq:Dirkernel_Atc}, for a random sample of size $n$ from $f_m$, is
\begin{align*}
    \mathrm{MISE}_{1,m}(h) = n^{-1} C_d(h)+\boldsymbol{\pi}^\top\big[(1-n^{-1})\boldsymbol{\Psi}_2(h)-2\boldsymbol{\Psi}_1(h)+\boldsymbol{\Psi}_0(h)\big]\boldsymbol{\pi},
\end{align*}
where 
\begin{align*}
   C_d(h) := \sum_{j=1}^m \pi_j \int_{\Delta^d} H\left(\frac{\yb}{h}+c\mathbf{1},\,\frac{\yb}{h}+c\mathbf{1}\right)
    \,\frac{C_\Dir(\yb/h + c\mathbf{1})^2}{C_\Dir(2\yb/h + 2c\mathbf{1} - \mathbf{1})}
    \, f_\Dir(\yb; \alphab_j)\,\lambda_d(\rd\yb), 
\end{align*}
and the $m\times m$ matrices $\boldsymbol{\Psi}_a(h)$, $a=0,1,2$, have entries:
\begin{align*}
    \boldsymbol{\Psi}_0(h) &:= \left(H\left(\alphab_i, \alphab_j\right)\,\frac{C_{\Dir}(\alphab_i)C_{\Dir}(\alphab_j)}{C_{\Dir}(\alphab_i+\alphab_j-\mathbf{1})}\right)_{ij},\\
    \boldsymbol{\Psi}_1(h) &:= \left(\int_{\Delta^d} H\left(\frac{\yb}{h} + c\mathbf{1}, \alphab_i\right)\,\frac{C_{\Dir}(\yb/h+c\mathbf{1})C_{\Dir}(\alphab_i)}{C_{\Dir}(\yb/h+\alphab_i+c\mathbf{1}-\mathbf{1})}\, f_\Dir(\yb;\alphab_j) \,\lambda_d(\rd\yb)\right)_{ij},\\
    \boldsymbol{\Psi}_2(h) &:= \bigg(\int_{\Delta^d}\int_{\Delta^d} H\left(\frac{\yb_1}{h}+c\mathbf{1},\,\frac{\yb_2}{h}+c\mathbf{1}\right)\,\frac{C_{\Dir}(\yb_1/h+c\mathbf{1})C_{\Dir}(\yb_2/h+c\mathbf{1})}{C_{\Dir}((\yb_1+\yb_2)/h+2c\mathbf{1}-\mathbf{1})}\\ &~~~~ \times f_\Dir(\yb_1;\alphab_i)f_\Dir(\yb_2;\alphab_j) \,\lambda_d(\rd\yb_1)\,\lambda_d(\rd\yb_2)\bigg)_{ij}.
\end{align*} 
\end{corollary}

Although $c=1$ is not admissible for $r=1$, formally setting $c=1$ recovers the general structure of Proposition~\ref{prp:exact_mise}, with an additional factor $H$.

\begin{figure}[!h]
    \centering
    \includegraphics[page=1, width=0.48\textwidth, trim=0.12cm 0.6cm 1cm 1.9cm, clip]{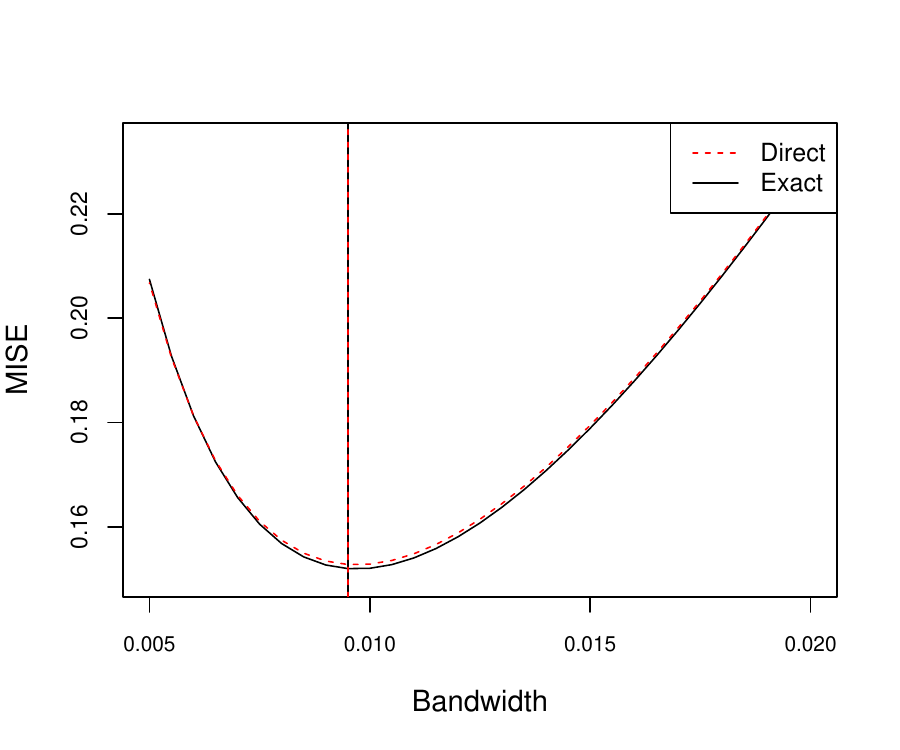}
    \hspace{0.02\textwidth}
    \includegraphics[page=2, width=0.48\textwidth, trim=0.12cm 0.6cm 1cm 1.9cm, clip]{img/MISE_curve.pdf}
    \caption{\small MISE curves obtained from the exact MISE representations (solid, black) and the direct Monte Carlo approximation (dashed, red) in density $(r=0)$ and projected first-derivative $(r=1)$ estimation, with vertical lines indicating the corresponding minimizers.}
    \label{fig:MISE_curve}
\end{figure}

To numerically assess the accuracy of the exact MISE representations in Proposition~\ref{prp:exact_mise} and Corollary~\ref{coro:exact_mise_r1}, we carry out a small numerical experiment on the Dirichlet mixture $\tfrac{1}{2} \,\Dir(5, 7, 11)\,+\,\tfrac{1}{2} \, \Dir(11, 7, 5)$ with sample size $n=500$. This mixture is also used as test case~(2) in the simulation study, as shown in Figure~\ref{fig:simumix} and Table~\ref{tab:para_mix}. The MISE was evaluated over a bandwidth grid from $0.005$ to $0.02$, with step size $5 \times 10^{-4}$. A direct Monte Carlo approximation was used for verification. In particular, we generated $B$ training samples $\mathcal{T}_1, \dots, \mathcal{T}_B$, each of size $n$, where $\mathcal{T}_b = \{\Xb_{b1}, \ldots, \Xb_{bn}\}$ and $\Xb_{b1}, \ldots, \Xb_{bn} \overset{\mathrm{iid}}{\sim} f$, with $f$ denoting the target density. We also used a common set of evaluation points
$\boldsymbol{s}_1,\ldots,\boldsymbol{s}_Q \overset{\mathrm{iid}}{\sim} \Dir(\oneb)$. The MISE of the bandwidth $h$ is then approximated by
\begin{align}
    \widehat{\mathrm{MISE}}_r(h) =
    \frac{1}{BQ\Gamma(D)}
    \sum_{b=1}^{B}
    \sum_{q=1}^{Q}
    \,\Bigl\|
    \widehat{\tDsf^{\otimes r} f}_h(\boldsymbol{s}_q;\mathcal{T}_b)
    - \tDsf^{\otimes r} f(\boldsymbol{s}_q)
    \Bigr\|^2.
    \label{eq:MCmise}
\end{align}
We set $B = 10^3$ and $Q = 10^4$. Figure~\ref{fig:MISE_curve} compares the resulting MISE curves in density and projected first-derivative estimation, with the vertical lines indicating the corresponding minimizers. In both panels, the two curves are nearly identical. Minor local discrepancies or a vertical shift may nevertheless arise from Monte Carlo error but have little effect on the location of the minimum.

The usefulness of the exact MISE representation is analyzed with a variant of the previous numerical experiment. Table~\ref{tab:time} reports the average computation times and standard errors of the two MISE evaluation methods for density and projected first-derivative estimation across different sample sizes $n$, based on $100$ repetitions. We set $B = 10^2$, $Q = 10^3$, and $S = 5 \times 10^5$, where $S$ denotes the Monte Carlo sample size used for each Dirichlet component in the exact MISE representations. Except for $r = 0, n = 100$, evaluation based on the exact MISE representations is substantially faster than the direct Monte Carlo approximation, with the difference being more pronounced for $r=1$. As $n$ increases, the computation times and standard errors of the exact MISE representation remain relatively stable. In contrast, the direct Monte Carlo approximation becomes up to to $6.5$ and $8.6$ times more expensive and its standard errors remain larger than those of the exact MISE representations.

\begin{table}[!h]
    \centering
    \caption{\small Average computation times (in seconds; left) and standard errors (right) for the exact MISE representation and direct Monte Carlo approximation in density $(r=0)$ and projected first-derivative $(r=1)$ estimation across different sample sizes $n$, based on 100 repetitions.}
    \label{tab:time}
    \begin{minipage}[t]{0.45\textwidth}
        \centering
        \begin{tabular}{clrrrr}
            \toprule
            $r$ & Method & 100 & 300 & 500 & 700 \\
            \midrule
            \multirow[c]{2}{*}{0} &
            Exact  & 1.92 & 1.89 & 2.10 & 1.84 \\
            & Direct & 1.72 & 5.17 & 11.68 & 12.04 \\
            \midrule
            \multirow[c]{2}{*}{1} &
            Exact  & 2.25 & 2.36 & 2.64 & 2.57 \\
            & Direct & 3.12 & 9.19 & 17.56 & 22.15 \\
            \bottomrule
        \end{tabular}
    \end{minipage}
    \hspace{0.5cm}
    \begin{minipage}[t]{0.46\textwidth}
        \centering
        \begin{tabular}{clrrrr}
            \toprule
            $r$ & Method & 100 & 300 & 500 & 700 \\
            \midrule
            \multirow[c]{2}{*}{\makecell[c]{0 \\$^{(\times 10^{-3})}$}} &
            Exact  & 0.65 & 0.68 & 0.67 & 0.68 \\
            & Direct & 2.61 & 1.25 & 0.96 & 0.71 \\
            \midrule
            \multirow[c]{2}{*}{\makecell[c]{1 \\$^{(\times 0.1)}$}} &
            Exact  & 1.23 & 1.31 & 1.20 & 1.11 \\
            & Direct & 4.90 & 2.76 & 2.19 & 1.58 \\
            \bottomrule
        \end{tabular}
    \end{minipage}
\end{table}

\subsection{Cross-validation}

Since LSCV is an unbiased estimator of the data-dependent part of the ISE, we begin with the ISE corresponding to \eqref{eq:kde_derivative}, namely
\begin{align*}
    \mathrm{ISE}_r(h) &= \int_{\Delta^d}\big\|\widehat{\tDsf^{\otimes r} f}_h(\xb) - \tDsf^{\otimes r} f(\xb)\big\|^2 \,\lambda_d(\rd\xb) \\
    &= \int_{\Delta^d} \big\|\widehat{\tDsf^{\otimes r} f}_h(\xb)\big\|^2 \,\lambda_d(\rd\xb) - 2\int_{\Delta^d} \widehat{\tDsf^{\otimes r} f}_h(\xb)^\top \tDsf^{\otimes r} f(\xb) \,\lambda_d(\rd\xb) + \int_{\Delta^d} \big\|\tDsf^{\otimes r} f(\xb)\big\|^2 \,\lambda_d(\rd\xb).
\end{align*}
For the projected first derivative, the data-dependent part is
\begin{align}
    \int_{\Delta^d} \big\|\widehat{\tDsf f}_h(\xb)\big\|^2 \,\lambda_d(\rd\xb) - 2\int_{\Delta^d} \widehat{\tDsf f}_h(\xb)^\top \tDsf f(\xb) \,\lambda_d(\rd\xb). \label{eq:ISE_r1_data_dependent}
\end{align}

To evaluate the cross-product term in \eqref{eq:ISE_r1_data_dependent} without calculating $\tDsf f$, we use integration by parts on the simplex to transfer the derivative from $f$ to the estimator, yielding an expectation with respect to $f$.

\begin{proposition} \label{prp:int_by_parts}
  Let $f$ be $C^1$ and $g$ be $C^2$ on an open set containing $\Delta^d$, such that
  \begin{enumerate}[label=(\roman*)]
    \item \label{prp:int_by_parts_1} $\|\tDsf f\| \, \|\tDsf g\|$ and $|f| \, \|\tDsf^{\otimes 2} g\|$ belong to $L^1(\Delta^d)$;
    \item \label{prp:int_by_parts_2} $f(\xb) \, \tDsf g(\xb) \to \mathbf{0}$ as $\xb \to \xb^*$ within $\Delta^d$, for every $\xb^* \in \partial\Delta^d$.
  \end{enumerate}
  Then
  \begin{align}
    \int_{\Delta^d} \tDsf f(\xb)^\top \tDsf g(\xb) \,\lambda_d(\rd\xb) = -\vecop(\Pb)^\top \int_{\Delta^d} \Dsf^{\otimes 2} g(\xb) f(\xb) \,\lambda_d(\rd\xb). \label{eq:int_by_parts}
  \end{align}
  In particular, \eqref{eq:int_by_parts} holds with $g = \hat{f}_h$ if $c \geq 2$, $f$ is bounded on $\Delta^d$, and $\|\tDsf f\| \in L^1(\Delta^d)$.
\end{proposition}

Hence, if $c \geq 2$ and the density $f$ is as in the last claim of Proposition~\ref{prp:int_by_parts}, then
\begin{align*}
  \int_{\Delta^d} \widehat{\tDsf f}_h(\xb)^\top \tDsf f(\xb) \,\lambda_d(\rd\xb) = -\vecop(\Pb)^\top \int_{\Delta^d} \widehat{\Dsf^{\otimes2} f}_h(\xb) f(\xb)\,\lambda_d(\rd\xb) = -\vecop(\Pb)^\top \mathbb{E}_{\Xb \sim f}\big[\Dsf^{\otimes 2} \hat{f}_h(\Xb)\big].
\end{align*}
Replacing this expectation by the leave-one-out average yields the LSCV for the projected first derivative:
\begin{align}
    \mathrm{LSCV}_1(h) = \int_{\Delta^d} \big\|\widehat{\tDsf f}_h(\xb)\big\|^2 \,\lambda_d(\rd\xb) + \frac{2}{n}\vecop(\Pb)^\top \sum_{i=1}^n \Dsf^{\otimes 2} \hat{f}_h^{-i}(\Xb_i).
    \label{eq:LSCV_r1}
\end{align}
Based on \eqref{eq:ISE_r1_data_dependent}, the RSLS is
\begin{align}
    \mathrm{RSLS}_1(h) &= \int_{\Delta^d} \bigl\|\,\widehat{\tDsf f}_h(\xb)\bigr\|^2 \,\lambda_d(\rd\xb) - 2\int_{\Delta^d} \widehat{\tDsf f}_h(\xb)^\top \tDsf f_m(\xb)\,\lambda_d(\rd\xb).
    \label{eq:RSLS_r1}
\end{align}

Criteria \eqref{eq:LSCV_r1} and \eqref{eq:RSLS_r1} admit the following explicit forms, which avoid numerical integration.

\begin{corollary} \label{coro:lscv_r1}
    Let $f$ satisfy the conditions of Proposition~\ref{prp:int_by_parts}, and let $c \geq 2$. Then,
    \begin{align*}
        \mathrm{LSCV}_1(h) &= \frac{1}{n^2} \sum_{i,\,j=1}^n H_{ij}\, \frac{C_{\Dir}(\Xb_i/h+c\oneb)C_{\Dir}(\Xb_j/h+c\oneb)}{C_{\Dir}((\Xb_i+\Xb_j)/h + 2c\oneb - \oneb)} + \frac{2}{n}\vecop(\Pb)^\top \sum_{i=1}^n \Dsf^{\otimes 2} \hat{f}_h^{-i}(\Xb_i)\\
        &= \frac{\Gamma(1/h+Dc)}{n^2}\sum_{i,j=1}^n \bigg\{
        \frac{\Gamma(1/h+Dc)}{\Gamma(2/h+2Dc-
        D)}\,H_{ij}\, \prod_{k=1}^D \frac{\Gamma((X_{ik}+X_{jk})/h+2c-1)}{\Gamma(X_{ik}/h+c)\Gamma(X_{jk}/h+c)}\nonumber\\
        &~~~~ +2(1-n^{-1})^{-1}(1-\delta_{ij})\Big(\sum_{s,t=1}^D(\delta_{st}-D^{-1}) M_{ij,st}\Big) \Big(\prod_{k=1}^D \frac{X_{ik}^{X_{jk}/h+c-1}}{\Gamma(X_{jk}/h+c)}\Big)\bigg\},
    \end{align*}
    where $\boldsymbol{M}_{ij}$ is a $D\times D$ matrix with $(s,t)$th entries
    \begin{align*}
        M_{ij,st} = (X_{is}X_{it})^{-1}(X_{js}/h+c-1)(X_{jt}/h+c-1-\delta_{st}),
    \end{align*}
    and $H_{ij} := H(\Xb_i/h+c\oneb, \Xb_j/h+c\oneb)$, with $H$ defined in \eqref{eq:Heq}.
\end{corollary}

\begin{corollary} \label{coro:rsls_r1}
    Let $f_m$ be the $m$-component Dirichlet mixture in \eqref{eq:fm} with $\alphab_j>\tfrac32\oneb$, and let $c \geq \tfrac32$. Then,
    \begin{align*}
        \mathrm{RSLS}_1(h) 
        &= \frac{1}{n^2} \sum_{i,\,j=1}^n H_{ij}\, \frac{C_{\Dir}(\Xb_i/h+c\oneb)C_{\Dir}(\Xb_j/h+c\oneb)}{C_{\Dir}((\Xb_i+\Xb_j)/h + 2c\oneb - \oneb)} \\&~~~~ - \frac{2}{n}\sum_{s=1}^n\sum_{t=1}^m G_{st}\,\pi_t \frac{C_\Dir(\Xb_s/h+c\oneb)C_\Dir(\alphab_t)}{C_\Dir(\Xb_s/h + \alphab_t + c\oneb - \oneb)},
    \end{align*}
    where $H_{ij} := H(\Xb_i/h+c\oneb, \Xb_j/h+c\oneb)$ and $G_{st} := H(\Xb_s/h+c\oneb, \alphab_t)$, with $H$ defined in \eqref{eq:Heq}.
\end{corollary}

As above, formally setting $c=1$ in Corollaries~\ref{coro:lscv_r1} and~\ref{coro:rsls_r1} recovers the structure of Proposition~\ref{prp:lscv} and Corollary~\ref{coro:rsls}, respectively, apart from the $H$ factors. 

Compared with density estimation, density derivative estimation is inherently more variable \citep{Chacon2018}. This additional variability further increases the instability of the LSCV in the projected first derivative setting, while RSLS is expected to be more stable. To illustrate this, we followed the setting of Figure~\ref{fig:MISE_curve} and generated $10^3$ independent samples. For each sample, we evaluated the LSCV and RSLS, together with an oracle reference-smoothed least-squares criterion (ORSLS), obtained by using the true target density as the reference density in RSLS. ORSLS is unbiased for the MISE up to an additive constant. It serves as an oracle benchmark for isolating the effect of reference density estimation and the variability induced by the leave-one-out average. 

\begin{figure}[!tb]
    \centering
    \includegraphics[page=1, width=0.48\textwidth, trim=0.1cm 0.6cm 1cm 2cm, clip]{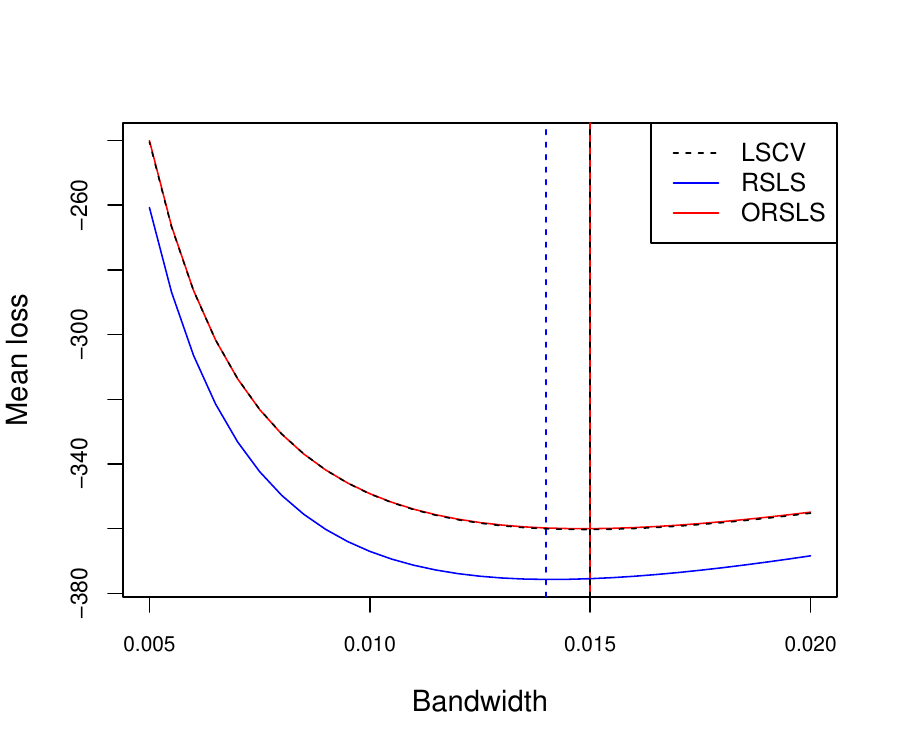}
    \hspace{0.02\textwidth}
    \includegraphics[page=2, width=0.48\textwidth, trim=0.1cm 0.6cm 1cm 2cm, clip]{img/CV_explore.pdf}
    \caption{\small Criterion curves with the vertical lines indicating the corresponding minimizers (left) and distributions of the selected bandwidths (right) by LSCV, RSLS, and ORSLS for projected first derivatives.}
    \label{fig:cv_explore}
\end{figure}

The left panel of Figure~\ref{fig:cv_explore} presents the three criterion curves averaged over the $10^3$ samples, with the vertical lines indicating the minimizers of the corresponding mean curves. Accordingly, the mean curve of LSCV is nearly identical to that of ORSLS. Moreover, its minimizer coincides with those obtained from Corollary~\ref{coro:exact_mise_r1} and direct Monte Carlo approximation, as shown in the right panel of Figure~\ref{fig:MISE_curve}. The mean RSLS curve has a similar shape, but its level and minimizer are shifted because of the bias introduced by the estimated reference density. The right panel presents the distributions of the selected bandwidths. The bandwidths selected by RSLS and ORSLS are concentrated near the minimizer of their mean curves. Although the RSLS distribution is slightly shifted relative to the that of ORSLS, the two have similar shapes and overlap substantially. By contrast, the bandwidths selected by LSCV are widely dispersed. Since the search is restricted to a finite interval, many selections accumulate at the endpoints when the optimal bandwidths lie outside the search range, illustrating the instability of LSCV bandwidth selection.

\section{Simulation studies}
\label{sec:simu}

In this section, we compare the performance of the three bandwidth selectors, $\hat{h}_\mathrm{EMI}$, $\hat{h}_\mathrm{LSCV}$, and $\hat{h}_\mathrm{RSLS}$ for density and first-derivative estimation. For density estimation, we additionally include the selector $\hat{h}_{\mathrm{LCV}}$ introduced by \cite{Aitchison1985}. As an oracle benchmark, we report results using a reference bandwidth minimizing the MISE, denoted by $h_{\mathrm{MISE}}$, computed from the known target density $f$ of each test mixture. Its MISE serves as an estimate of the best performance achievable with a single fixed bandwidth. For Dirichlet mixture targets, $h_{\mathrm{MISE}}$ can be obtained directly by minimizing the exact MISE in Proposition~\ref{prp:exact_mise} and Corollary~\ref{coro:exact_mise_r1} for density and first-derivative estimation, respectively. For other targets, $h_{\mathrm{MISE}}$ is obtained by direct Monte Carlo approximation in \eqref{eq:MCmise} with $B = 10^3$ and $Q = 10^4$. The bandwidths were chosen over a grid $h \in [10^{-3}, 1]$.

\begin{figure}[!t]
  \centering
  \foreach \p [count=\i from 1] in {1,2,3,4,5,6,7,8,9,10,11,12}{
    \begin{overpic}[page=\p,width=0.24\linewidth,
      clip,trim={1.3cm 1.4cm 1.2cm 2cm}]{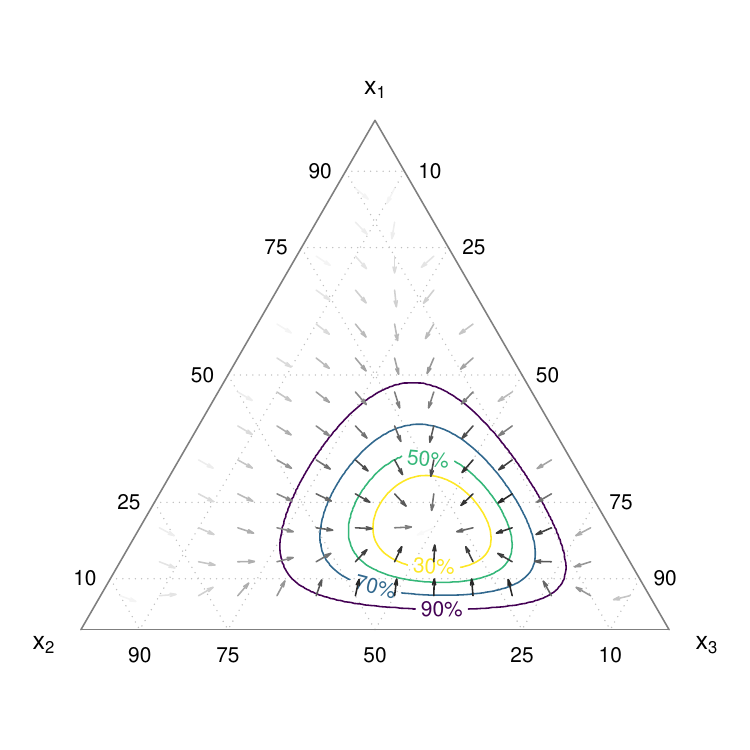}
      \put(44,-9){\footnotesize (\i)}%
    \end{overpic}%
    \ifnum\i=4\par\vspace{15pt}\fi
    \ifnum\i=8\par\vspace{15pt}\fi
  }%
  \vspace{0.5cm}
  \caption{\small Probability contours and gradient fields for the test mixtures on $\Delta^2$. The projected gradient vector fields are unit-normalized, with higher transparency indicating lower norm.}
  \label{fig:simumix}
\end{figure}

The selectors are evaluated on $\Delta^2$ and $\Delta^4$. For each dimension, six Dirichlet mixtures and six Gaussian mixtures are constructed as test cases, with sample size $n = 100$ and $500$ considered. The Gaussian mixtures and the generated samples are mapped onto the simplex using the inverse ilr transformation. Figure~\ref{fig:simumix} displays the probability contours and gradient fields of the test mixtures on $\Delta^2$, and the corresponding mixture parameters are provided in Table~\ref{tab:para_mix}. For each Dirichlet mixture component, $\alphab = \thetab/h + \oneb$, where $\thetab$ denotes the mode and $h$ is the scale. For each Gaussian-ilr mixture component, $\mub$ denotes the mean and $\operatorname{vech} \Sigmab$ denotes the half-vectorization of the covariance matrix $\Sigmab$.

\renewcommand{\arraystretch}{1.5}
\begin{table}[!htb]
    \centering
    \caption{\small Parameters of the test mixtures in $\Delta^2$, with Gaussian mixtures mapped onto the simplex via the inverse ilr transformation.}
    \small
    \begin{tabular}{c
        >{\raggedright\hangindent=1em\arraybackslash}p{13cm}}
        \toprule
        \textbf{Dirichlet} & $\pi_1 \Dir(\thetab_1; h_1) + \cdots + \pi_m \Dir(\thetab_m; h_m)$ \\
        \addlinespace[0.3em]
        1 & $\Dir\bigl((0.2, 0.3, 0.5); \, 0.1\bigr)$ \\
        2 & $\frac{1}{2} \,\Dir\bigl((0.5, 0.3, 0.2); \, 0.05 \bigr) +  \frac{1}{2} \, \Dir\bigl((0.2, 0.3, 0.5); \, 0.05\bigr)$ \\
        3 & $ \frac{1}{4} \, \sum_{i=1}^4 \Dir\Bigl(\bigl(\frac{i}{6}, \frac{i}{6} (1 - \frac{i}{6}), (1 - \frac{i}{6})^2\bigr); \, 0.08\Bigr) $ \\
        4 & $\frac{1}{5} \, \Dir\bigl((\frac{1}{3}, \frac{31}{75}, \frac{19}{75}); \, 0.1 \bigr) + \frac{7}{15} \, \Dir\bigl((\frac{1}{3}, 0.1, \frac{17}{30}); \, 0.05\bigr) + \frac{4}{15} \, \Dir\left((\frac{2}{15}, \frac{77}{150}, \frac{53}{150}); \, 0.035\right) + \frac{1}{15} \, \Dir\bigl((0.44, 0.42, 0.14); \, 0.04\bigr)$ \\
        5 & $\frac{1}{6} \, \sum_{i=1}^3 \Dir\Bigl(\bigl(\frac{i}{5}, \frac{2}{9}(1 - \frac{i}{5}), \frac{7}{9}(1 - \frac{i}{5})\bigr); \, \frac{4-i}{50}\Bigr) +$ $\frac{1}{6} \, \sum_{i=1}^3 \Dir\Bigl(\bigl(\frac{7i}{60}, \frac{4}{5}(1 - \frac{7i}{60}), \frac{1}{5}(1 - \frac{7i}{60})\bigr); \, \frac{7(4-i)}{300}\Bigr)$ \\
        6 & $\frac{9}{100} \, \sum_{i=1}^5 \Dir\Bigl(\bigl(\frac{2i}{15}, \frac{2i}{15} (1 - \frac{2i}{15}), (1 - \frac{2i}{15})^2\bigr); \, \frac{9}{100}\sqrt{\frac{2i}{15}}\Bigr)$ + $\frac{11}{100} \, \sum_{i=1}^5 \Dir\Bigl(\bigl(\frac{2i}{15}(1 - \frac{2i}{15}) + \frac{4}{225}, \frac{2i}{15} - \frac{4}{225}, (1 - \frac{2i}{15})^2\bigr); \, \frac{1}{10}\sqrt{\frac{12-2i}{15}}\Bigr)$ \\  
        \midrule
        \textbf{Gaussian-ilr} & $\pi_1 \Norm(\mub_1; \operatorname{vech} \Sigmab_1) + \cdots + \pi_m \Norm(\mub_m; \operatorname{vech} \Sigmab_m)$ \\
        \addlinespace[0.3em]
        7 & $\Norm\bigl(\mathbf{0};\,(0.25,0,1)\bigr)$ \\
        8 & $\frac{1}{2}\,\Norm\bigl((0.75,0);\,\frac{4}{9}\mathbf{I}\bigr) + \frac{1}{2}\,\Norm\bigl((-0.75,0);\,\frac{4}{9}\mathbf{I}\bigr)$ \\
        9 & $\frac{1}{2}\,\Norm\bigl((0.4,-0.4);\,(\frac{4}{9}, \frac{14}{45}, \frac{4}{9})\bigr) + \frac{1}{2}\,\Norm\bigl((-0.4,0.4);\,\frac{4}{9}\mathbf{I}\bigr)$ \\
        10 & $\frac{3}{7}\,\Norm\bigl((-0.6, 0); (\frac{9}{25}, \frac{63}{250}, \frac{49}{100})\bigr) + \frac{3}{7}\,\Norm\bigl((0.5, 0.6); (\frac{9}{25}, 0, \frac{49}{100})\bigr) + \frac{1}{7}\,\Norm\bigl((0.5, -0.6); (\frac{9}{25}, 0, \frac{49}{100})\bigr)$ \\
        11 & $\frac{11}{25}\,\Norm\bigl((-1, 0); (\frac{4}{9}, \frac{4}{15}, \frac{4}{9})\bigr) + \frac{11}{25}\,\Norm\bigl((1, 0); (\frac{4}{9}, \frac{4}{15}, \frac{4}{9})\bigr) + \frac{3}{25}\,\Norm\bigl(\mathbf{0}; (\frac{1}{9}, \frac{1}{15}, \frac{1}{9})\bigr)$ \\
        12 & $\frac{1}{5}\Norm(\mathbf{0};\,\frac{1}{10}\mathbf{I}) + \frac{1}{5}\sum_{(i,j)\in \{-1,1\}^2}\Norm\bigl(\frac{4}{5}(i, j);\,\frac{1}{10}\mathbf{I}\bigr)$\\
        \bottomrule
    \end{tabular}
    \label{tab:para_mix}
\end{table}
\renewcommand{\arraystretch}{1}

To extend mixtures from $\Delta^2$ to $\Delta^d$, we follow the construction of \cite{Xie2025}. For the Dirichlet mixture, given a component mode $\thetab \in \Delta^2$, we define the corresponding mode as 
\begin{align*}
    \widetilde{\thetab} = \left(\frac{3}{D}\thetab^\top,\, \frac{1}{D}\oneb_{d-2}^\top \right)^\top \in\Delta^d.
\end{align*}
The scale $h$ and mixing proportions $\pib$ are kept unchanged. For a Gaussian mixture, the mean vector $\mub \in \Rb^2$ and covariance matrix $\Sigmab \in \Rb^{2\times2}$ in each component are extended as
\begin{align*}
    \widetilde{\mub} = \left(\mub^\top,\, \mathbf{0}_{d-2}^\top \right)^\top \in \Rb^d,
    ~~~~ \widetilde{\Sigmab} = 
    \begin{pmatrix}
        \Sigmab & \boldsymbol{0} \\
        \boldsymbol{0} &
        \operatorname{diag}(\nu_1,\ldots,\nu_{d-2})
    \end{pmatrix} \in \Rb^{d\times d}.
\end{align*}
Let $\lambda_1 \leq \lambda_2$ denote the two eigenvalues of $\Sigmab$. The additional variances are defined by
\begin{align*}
    \nu_k = \lambda_1 + \frac{k}{d-1}(\lambda_2 - \lambda_1),~~~
    k=1,\dots,d-2.
\end{align*}

The same random samples were used across the different selectors to ensure a fair comparison. Within each of $10^3$ Monte Carlo repetitions, the fitted estimators were evaluated at a common set of $10^4$ points drawn uniformly from the simplex, and the MISE and mean integrated absolute error (MIAE) were approximated by Monte Carlo. Tables~\ref{tab:MISE_r0} and~\ref{tab:MISE_r1} report the MISE results in the density and first-derivative simulation studies, respectively. In each scenario, boldface indicates the minimum among data-driven selectors and values not significantly different from it by a paired $t$-test across the $10^3$ ISEs at the $5\%$ level. The discussion below focuses on the MISE results, while the corresponding MIAE results, which lead to similar conclusions, are reported and discussed in Appendix~\ref{sec:simu_MIAE}.


\begin{sidewaystable}[p]
    \centering
    \caption{\small MISE values for the density estimation simulations, with Monte Carlo standard errors shown as subscripts. Boldface indicates the minimum and values not significantly different from it among the data-driven selectors in each scenario. The oracle benchmark is the MISE selector.}
    \label{tab:MISE_r0}
    
    \begin{tabular}{llcccccccccccc}
    \toprule
    Setting & Selector & \multicolumn{6}{c}{Dirichlet} & \multicolumn{6}{c}{Gaussian-ilr} \\
    \cmidrule(lr){3-8} \cmidrule(lr){9-14}
    & & (1) & (2) & (3) & (4) & (5) & (6) & (7) & (8) & (9) & (10) & (11) & (12) \\
    \midrule
    
    \multirow{5}{*}{\centering\makecell[l]{$d=2$, \\ $n=100$ \\ [3pt] ($\times 0.1$)}}
    
    & MISE & $2.44_{0.04}$ & $3.98_{0.04}$ & $2.11_{0.02}$ & $4.55_{0.05}$ & $5.43_{0.04}$ & $2.51_{0.02}$ & $2.25_{0.02}$ & $1.89_{0.02}$ & $2.11_{0.02}$ & $2.65_{0.02}$ & $4.37_{0.04}$ & $7.55_{0.05}$ \\
    & EMI & $2.54_{0.04}$ & $\mathbf{4.13}_{0.05}$ & $\mathbf{2.26}_{0.03}$ & $\mathbf{4.81}_{0.05}$ & $\mathbf{5.76}_{0.05}$ & $2.90_{0.03}$ & $\mathbf{2.44}_{0.03}$ & $2.43_{0.03}$ & $2.63_{0.01}$ & $3.70_{0.04}$ & $4.72_{0.04}$ & $8.34_{0.09}$ \\
    & LSCV & $3.22_{0.06}$ & $4.72_{0.07}$ & $2.53_{0.04}$ & $5.14_{0.06}$ & $6.02_{0.06}$ & $2.97_{0.04}$ & $2.63_{0.04}$ & $2.21_{0.04}$ & $2.37_{0.03}$ & $3.06_{0.04}$ & $4.88_{0.05}$ & $8.11_{0.06}$ \\
    & RSLS & $\mathbf{2.51}_{0.04}$ & $4.20_{0.04}$ & $2.33_{0.03}$ & $4.92_{0.05}$ & $5.99_{0.05}$ & $\mathbf{2.76}_{0.03}$ & $2.51_{0.03}$ & $\mathbf{2.07}_{0.03}$ & $\mathbf{2.20}_{0.02}$ & $2.94_{0.03}$ & $4.84_{0.04}$ & $8.52_{0.08}$ \\
    & LCV & $2.73_{0.04}$ & $4.29_{0.05}$ & $2.37_{0.03}$ & $4.91_{0.05}$ & $\mathbf{5.74}_{0.05}$ & $\mathbf{2.78}_{0.03}$ & $2.52_{0.03}$ & $2.14_{0.03}$ & $2.34_{0.02}$ & $\mathbf{2.90}_{0.03}$ & $\mathbf{4.64}_{0.04}$ & $\mathbf{7.88}_{0.06}$ \\

    \midrule

    \multirow{5}{*}{\centering\makecell[l]{$d=2$, \\ $n=500$ \\ [3pt] ($\times 0.1$)}}
    
    & MISE & $0.94_{0.01}$ & $1.52_{0.01}$ & $0.79_{0.01}$ & $1.70_{0.01}$ & $2.11_{0.01}$ & $1.00_{0.01}$ & $0.93_{0.01}$ & $0.76_{0.01}$ & $0.89_{0.01}$ & $1.07_{0.01}$ & $1.88_{0.01}$ & $2.89_{0.02}$ \\
    & EMI & $0.96_{0.01}$ & $1.54_{0.01}$ & $\mathbf{0.81}_{0.01}$ & $1.72_{0.01}$ & $\mathbf{2.14}_{0.01}$ & $\mathbf{1.03}_{0.01}$ & $\mathbf{0.96}_{0.01}$ & $\mathbf{0.78}_{0.01}$ & $1.18_{0.02}$ & $\mathbf{1.10}_{0.01}$ & $2.01_{0.01}$ & $\mathbf{2.94}_{0.02}$ \\
    & LSCV & $1.06_{0.01}$ & $1.63_{0.02}$ & $0.86_{0.01}$ & $1.81_{0.02}$ & $2.20_{0.02}$ & $1.07_{0.01}$ & $1.02_{0.01}$ & $0.82_{0.01}$ & $0.96_{0.01}$ & $1.14_{0.01}$ & $1.98_{0.01}$ & $2.98_{0.02}$ \\
    & RSLS & $\mathbf{0.94}_{0.01}$ & $\mathbf{1.54}_{0.01}$ & $0.82_{0.01}$ & $\mathbf{1.71}_{0.01}$ & $2.15_{0.01}$ & $1.08_{0.01}$ & $0.96_{0.01}$ & $0.78_{0.01}$ & $1.17_{0.01}$ & $1.13_{0.01}$ & $2.23_{0.02}$ & $3.00_{0.02}$ \\
    & LCV & $0.99_{0.01}$ & $1.58_{0.01}$ & $0.83_{0.01}$ & $1.80_{0.02}$ & $2.19_{0.02}$ & $\mathbf{1.04}_{0.01}$ & $1.00_{0.01}$ & $0.80_{0.01}$ & $\mathbf{0.93}_{0.01}$ & $1.12_{0.01}$ & $\mathbf{1.94}_{0.01}$ & $\mathbf{2.94}_{0.02}$ \\
    
    \midrule
    
    \multirow{5}{*}{\centering\makecell[l]{$d=4$, \\ $n=100$}}

    & MISE & $9.43_{0.08}$ & $22.41_{0.14}$ & $9.53_{0.06}$ & $25.00_{0.15}$ & $45.24_{0.24}$ & $15.50_{0.09}$ & $10.09_{0.04}$ & $9.16_{0.05}$ & $9.65_{0.04}$ & $12.34_{0.05}$ & $25.85_{0.09}$ & $117.87_{0.43}$ \\
    & EMI & $\mathbf{9.62}_{0.08}$ & $\mathbf{22.87}_{0.14}$ & $\mathbf{9.86}_{0.06}$ & $\mathbf{25.59}_{0.15}$ & $\mathbf{46.86}_{0.24}$ & $\mathbf{15.99}_{0.09}$ & $11.44_{0.06}$ & $9.87_{0.06}$ & $10.26_{0.04}$ & $13.74_{0.07}$ & $26.94_{0.10}$ & $\mathbf{118.54}_{0.45}$ \\
    & LSCV & $10.44_{0.10}$ & $24.17_{0.20}$ & $10.34_{0.08}$ & $26.37_{0.16}$ & $47.41_{0.27}$ & $16.68_{0.11}$ & $10.73_{0.08}$ & $9.71_{0.07}$ & $10.16_{0.06}$ & $13.01_{0.07}$ & $27.48_{0.33}$ & $121.48_{0.49}$ \\
    & RSLS & $9.64_{0.08}$ & $23.20_{0.14}$ & $10.00_{0.06}$ & $25.71_{0.15}$ & $47.38_{0.24}$ & $16.14_{0.09}$ & $\mathbf{10.43}_{0.05}$ & $\mathbf{9.50}_{0.05}$ & $\mathbf{9.80}_{0.04}$ & $13.01_{0.06}$ & $27.33_{0.11}$ & $119.39_{0.45}$ \\
    & LCV & $9.99_{0.09}$ & $23.28_{0.15}$ & $10.03_{0.07}$ & $26.40_{0.16}$ & $48.51_{0.25}$ & $16.17_{0.09}$ & $10.51_{0.05}$ & $9.57_{0.05}$ & $10.03_{0.05}$ & $\mathbf{12.76}_{0.06}$ & $\mathbf{26.68}_{0.10}$ & $119.76_{0.45}$ \\
        
    \midrule

    \multirow{5}{*}{\centering\makecell[l]{$d=4$, \\ $n=500$}}
    
    & MISE & $4.45_{0.03}$ & $11.05_{0.06}$ & $4.52_{0.03}$ & $12.74_{0.06}$ & $23.91_{0.11}$ & $7.93_{0.04}$ & $5.60_{0.02}$ & $4.87_{0.02}$ & $5.44_{0.02}$ & $6.84_{0.02}$ & $15.42_{0.04}$ & $63.08_{0.19}$ \\
    & EMI & $4.49_{0.03}$ & $\mathbf{11.11}_{0.05}$ & $4.55_{0.02}$ & $\mathbf{12.83}_{0.06}$ & $\mathbf{24.10}_{0.11}$ & $\mathbf{7.99}_{0.04}$ & $\mathbf{5.68}_{0.02}$ & $\mathbf{4.94}_{0.02}$ & $6.27_{0.03}$ & $7.05_{0.03}$ & $16.60_{0.05}$ & $\mathbf{63.73}_{0.20}$ \\
    & LSCV & $4.64_{0.03}$ & $11.36_{0.06}$ & $4.69_{0.03}$ & $13.07_{0.06}$ & $24.40_{0.11}$ & $8.16_{0.04}$ & $5.75_{0.03}$ & $4.99_{0.02}$ & $5.56_{0.02}$ & $6.98_{0.02}$ & $\mathbf{15.68}_{0.05}$ & $63.81_{0.20}$ \\
    & RSLS & $\mathbf{4.47}_{0.03}$ & $11.12_{0.05}$ & $\mathbf{4.55}_{0.02}$ & $12.85_{0.06}$ & $24.22_{0.10}$ & $8.01_{0.04}$ & $5.90_{0.02}$ & $5.00_{0.02}$ & $6.90_{0.03}$ & $7.35_{0.03}$ & $17.74_{0.06}$ & $64.07_{0.21}$ \\
    & LCV & $4.61_{0.03}$ & $11.38_{0.06}$ & $4.64_{0.03}$ & $13.92_{0.07}$ & $27.21_{0.12}$ & $8.25_{0.04}$ & $\mathbf{5.67}_{0.02}$ & $\mathbf{4.94}_{0.02}$ & $\mathbf{5.50}_{0.02}$ & $\mathbf{6.91}_{0.02}$ & $16.04_{0.05}$ & $64.61_{0.20}$ \\
    
    \bottomrule
    \end{tabular}
\end{sidewaystable}


\begin{sidewaystable}[p]
    \centering
    \caption{\small MISE values for the first-derivative estimation simulations, with Monte Carlo standard errors shown as subscripts. Boldface indicates the minimum and values not significantly different from it among the data-driven selectors in each scenario. The oracle benchmark is the MISE selector.}
    \label{tab:MISE_r1}
    
    \begin{tabular}{llcccccccccccc}
    \toprule
    Setting & Selector & \multicolumn{6}{c}{Dirichlet} & \multicolumn{6}{c}{Gaussian-ilr} \\
    \cmidrule(lr){3-8} \cmidrule(lr){9-14}
    & & (1) & (2) & (3) & (4) & (5) & (6) & (7) & (8) & (9) & (10) & (11) & (12) \\
    \midrule
    
    \multirow{4}{*}{\centering\makecell[l]{$d=2$, \\ $n=100$ \\ [3pt] ($\times 10$)}}
    
    & MISE & $3.01_{0.04}$ & $8.55_{0.08}$ & $3.50_{0.03}$ & $13.58_{0.12}$ & $22.32_{0.15}$ & $6.07_{0.04}$ & $14.20_{0.07}$ & $6.71_{0.05}$ & $6.45_{0.04}$ & $11.94_{0.06}$ & $47.19_{0.23}$ & $48.68_{0.27}$ \\
    & EMI & $3.15_{0.04}$ & $\mathbf{8.91}_{0.09}$ & $\mathbf{3.90}_{0.05}$ & $\mathbf{14.46}_{0.14}$ & $\mathbf{24.01}_{0.17}$ & $7.61_{0.10}$ & $\mathbf{15.34}_{0.09}$ & $10.14_{0.12}$ & $9.60_{0.03}$ & $17.74_{0.17}$ & $55.53_{0.40}$ & $\mathbf{53.75}_{0.49}$ \\
    & LSCV & $41.66_{6.42}$ & $43.97_{5.06}$ & $22.90_{4.93}$ & $71.65_{9.45}$ & $65.25_{7.41}$ & $31.78_{5.23}$ & $69.74_{13.07}$ & $25.50_{5.92}$ & $20.95_{3.79}$ & $28.97_{3.16}$ & $182.48_{23.03}$ & $86.77_{5.50}$ \\
    & RSLS & $\mathbf{3.12}_{0.04}$ & $9.03_{0.08}$ & $4.00_{0.06}$ & $14.83_{0.14}$ & $25.01_{0.19}$ & $\mathbf{6.93}_{0.07}$ & $15.65_{0.12}$ & $\mathbf{7.39}_{0.07}$ & $\mathbf{6.92}_{0.04}$ & $\mathbf{13.47}_{0.07}$ & $\mathbf{53.46}_{0.27}$ & $54.60_{0.43}$ \\

    \midrule

    \multirow{4}{*}{\centering\makecell[l]{$d=2$, \\ $n=500$ \\ [3pt] ($\times 10$)}}
    
    & MISE & $1.43_{0.02}$ & $4.24_{0.04}$ & $1.74_{0.01}$ & $6.75_{0.05}$ & $12.25_{0.08}$ & $3.19_{0.02}$ & $9.44_{0.05}$ & $3.76_{0.02}$ & $3.81_{0.02}$ & $7.10_{0.03}$ & $30.16_{0.16}$ & $26.15_{0.14}$ \\
    & EMI & $1.47_{0.02}$ & $4.29_{0.04}$ & $\mathbf{1.79}_{0.01}$ & $6.80_{0.05}$ & $\mathbf{12.44}_{0.07}$ & $\mathbf{3.33}_{0.02}$ & $\mathbf{10.03}_{0.05}$ & $\mathbf{4.17}_{0.04}$ & $5.76_{0.08}$ & $\mathbf{7.59}_{0.04}$ & $\mathbf{35.69}_{0.28}$ & $\mathbf{28.03}_{0.14}$ \\
    & LSCV & $8.63_{1.04}$ & $10.82_{0.90}$ & $5.62_{0.51}$ & $18.28_{1.45}$ & $17.92_{0.76}$ & $5.74_{0.35}$ & $18.37_{1.76}$ & $7.06_{0.62}$ & $5.59_{0.20}$ & $11.34_{0.63}$ & $47.79_{2.38}$ & $34.13_{1.00}$ \\
    & RSLS & $\mathbf{1.43}_{0.02}$ & $\mathbf{4.27}_{0.04}$ & $1.82_{0.01}$ & $\mathbf{6.76}_{0.05}$ & $12.47_{0.07}$ & $3.42_{0.03}$ & $10.26_{0.05}$ & $4.34_{0.03}$ & $\mathbf{5.11}_{0.03}$ & $7.96_{0.05}$ & $38.83_{0.22}$ & $29.44_{0.15}$ \\
    
    \midrule
    
    \multirow{4}{*}{\centering\makecell[l]{$d=4$, \\ $n=100$ \\ [3pt] ($\times 10^{3}$)}}

    & MISE & $2.85_{0.02}$ & $9.83_{0.05}$ & $4.37_{0.03}$ & $19.17_{0.11}$ & $48.60_{0.22}$ & $8.91_{0.04}$ & $13.58_{0.03}$ & $9.12_{0.03}$ & $8.96_{0.02}$ & $14.54_{0.04}$ & $69.26_{0.13}$ & $250.23_{0.73}$ \\
    & EMI & $2.95_{0.02}$ & $\mathbf{10.16}_{0.06}$ & $\mathbf{4.63}_{0.03}$ & $\mathbf{19.74}_{0.12}$ & $\mathbf{50.64}_{0.23}$ & $\mathbf{9.34}_{0.05}$ & $16.50_{0.13}$ & $10.87_{0.08}$ & $10.57_{0.02}$ & $17.68_{0.10}$ & $\mathbf{72.89}_{0.26}$ & $253.21_{0.74}$ \\
    & LSCV & $4.46_{0.20}$ & $13.74_{0.42}$ & $9.75_{3.22}$ & $40.55_{17.01}$ & $55.03_{0.93}$ & $12.53_{1.16}$ & $16.92_{1.26}$ & $10.80_{0.27}$ & $9.85_{0.09}$ & $15.99_{0.16}$ & $93.36_{10.05}$ & $270.76_{2.27}$ \\
    & RSLS & $\mathbf{2.94}_{0.02}$ & $10.39_{0.11}$ & $4.67_{0.03}$ & $\mathbf{19.76}_{0.11}$ & $51.14_{0.23}$ & $\mathbf{9.33}_{0.05}$ & $\mathbf{14.35}_{0.17}$ & $\mathbf{9.47}_{0.05}$ & $\mathbf{9.06}_{0.02}$ & $\mathbf{15.46}_{0.09}$ & $\mathbf{72.76}_{0.21}$ & $\mathbf{252.79}_{0.75}$ \\
        
    \midrule

    \multirow{4}{*}{\centering\makecell[l]{$d=4$, \\ $n=500$ \\ [3pt] ($\times 10^{3}$)}}
    
    & MISE & $1.57_{0.01}$ & $5.62_{0.03}$ & $2.73_{0.01}$ & $11.87_{0.05}$ & $31.46_{0.12}$ & $5.50_{0.02}$ & $10.04_{0.02}$ & $6.13_{0.02}$ & $6.44_{0.01}$ & $10.40_{0.02}$ & $55.29_{0.12}$ & $166.87_{0.42}$ \\
    & EMI & $1.59_{0.01}$ & $\mathbf{5.65}_{0.02}$ & $\mathbf{2.76}_{0.01}$ & $\mathbf{11.94}_{0.05}$ & $\mathbf{31.63}_{0.13}$ & $\mathbf{5.54}_{0.02}$ & $\mathbf{10.59}_{0.03}$ & $6.47_{0.02}$ & $8.87_{0.05}$ & $11.19_{0.03}$ & $60.80_{0.20}$ & $\mathbf{173.80}_{0.43}$ \\
    & LSCV & $1.93_{0.03}$ & $6.31_{0.09}$ & $11.15_{7.91}$ & $29.31_{16.16}$ & $32.86_{0.15}$ & $6.01_{0.06}$ & $12.30_{1.10}$ & $\mathbf{6.37}_{0.02}$ & $\mathbf{6.67}_{0.02}$ & $\mathbf{10.82}_{0.06}$ & $\mathbf{58.71}_{0.64}$ & $\mathbf{172.78}_{0.73}$ \\
    & RSLS & $\mathbf{1.58}_{0.01}$ & $5.66_{0.02}$ & $\mathbf{2.76}_{0.01}$ & $11.95_{0.05}$ & $31.69_{0.13}$ & $5.56_{0.02}$ & $11.09_{0.03}$ & $6.85_{0.02}$ & $7.84_{0.02}$ & $11.83_{0.03}$ & $64.12_{0.15}$ & $174.83_{0.43}$ \\
    
    \bottomrule
    \end{tabular}
\end{sidewaystable}

\begin{table}[!t]
    \centering
    \caption{\small Ranking scores for selectors across different dimensions $d$ and sample sizes $n$ in density and first-derivative estimation. Scores are reported in the form $\text{Total}_{(\text{Dirichlet},\,\text{Gaussian-ilr})}$. A higher score indicates better performance. Boldface indicates the best-performing selector.}
    \label{tab:rank}
    \begin{tabular}{ccc|@{\hspace{14pt}}cccc}
    \toprule
    $r$ & $d$ & $n$ & EMI & LSCV & RSLS & LCV \\
    \midrule
    \multirow[t]{4}{*}{0}
    & \multirow[t]{2}{*}{2}
    & 100 & $7.79_{(4.96,\,2.83)}$ & $3.95_{(1.34,\,2.60)}$ & $8.61_{(4.43,\,4.18)}$ & $\mathbf{8.81}_{(3.90,\,4.91)}$ \\
    & & 500 & $\mathbf{9.91}_{(5.23,\,4.68)}$ & $4.32_{(1.65,\,2.67)}$ & $7.57_{(4.72,\,2.84)}$ & $7.61_{(3.16,\,4.45)}$ \\
    & \multirow[t]{2}{*}{4}
    & 100 & $8.68_{(6.00,\,2.68)}$ & $4.36_{(1.92,\,2.43)}$ & $\mathbf{9.43}_{(4.46,\,4.97)}$ & $7.16_{(2.44,\,4.72)}$ \\
    & & 500 & $\mathbf{9.65}_{(5.50,\,4.15)}$ & $6.67_{(2.45,\,4.22)}$ & $6.63_{(4.99,\,1.64)}$ & $6.64_{(1.91,\,4.73)}$ \\
    \midrule
    \multirow[t]{4}{*}{1}
    & \multirow[t]{2}{*}{2}
    & 100 & $9.38_{(5.27,\,4.11)}$ & $1.15_{(0.41,\,0.74)}$ & $\mathbf{9.91}_{(4.60,\,5.31)}$ & -- \\
    & & 500 & $\mathbf{10.27}_{(4.97,\,5.30)}$ & $2.58_{(0.84,\,1.73)}$ & $9.14_{(4.97,\,4.17)}$ & -- \\
    & \multirow[t]{2}{*}{4}
    & 100 & $8.11_{(5.33,\,2.78)}$ & $4.04_{(1.34,\,2.70)}$ & $\mathbf{10.64}_{(4.64,\,6.00)}$ & -- \\
    & & 500 & $\mathbf{9.19}_{(5.33,\,3.86)}$ & $6.71_{(1.42,\,5.29)}$ & $7.11_{(4.66,\,2.45)}$ & -- \\
    \bottomrule
    \end{tabular}
\end{table}

For density estimation, $\hat{h}_{\mathrm{EMI}}$ performs substantially better under the Dirichlet scenarios than under the Gaussian-ilr scenarios, mainly because its reference family is better matched to the Dirichlet targets. Its performance is also affected by the sample size $n$, particularly under the Gaussian-ilr scenarios, where performance markedly increases from $n=100$ to $n=500$. This is likely because the larger sample size reduces the uncertainty in reference mixture estimation. With respect to the comparison of $\hat{h}_{\mathrm{RSLS}}$ versus $\hat{h}_{\mathrm{LSCV}}$, the former outperforms the latter in all the Dirichlet scenarios, where the reference family is well matched to the target density, whereas, as $n$ increases for the Gaussian-ilr scenarios, $\hat{h}_{\mathrm{LSCV}}$ caches up with or surpasses $\hat{h}_{\mathrm{RSLS}}$, likely because the leave-one-out average becomes more stable and bias shows up in $\hat{h}_{\mathrm{RSLS}}$. This is also reflected in the standard errors of $\hat{h}_{\mathrm{LSCV}}$: they are generally larger than other selectors when $n = 100$, whereas the difference decreases for $n = 500$. Interestingly, $\hat{h}_{\mathrm{LCV}}$ achieves the best overall performance under the Gaussian-ilr scenarios. A possible explanation is that $\log \hat{f}_h^{-i}(\Xb_i) \to -\infty$ as $h \to 0$. This strongly penalizes overly small bandwidths, leading to more stable bandwidth selection than LSCV.

For first-derivative estimation, the performance of $\hat{h}_{\mathrm{EMI}}$ and $\hat{h}_{\mathrm{RSLS}}$ is broadly consistent with that observed for density estimation. In contrast, the instability of $\hat{h}_{\mathrm{LSCV}}$ is amplified, leading to generally poorer performance and substantially larger standard errors than those of the other selectors. However, this instability is reduced under the Gaussian-ilr scenarios with $d=4, n=500$, where $\hat{h}_{\mathrm{LSCV}}$ achieves the best performance. In addition to the effects of the reference density and sample size discussed above, the increase in dimension may also contribute to this improvement. As observations become more dispersed in $\Delta^4$, extremely large contributions may occur less frequently, thereby improving the stability.

To provide an overall comparison of the selectors in different scenarios, the ranking score is computed according to the criterion used in \cite{GarciaPortugues2013}: for each case, the $m$ bandwidth selectors $h_1, \dots,h_m$ are first sorted in increasing order of estimation error and assigned ranking points $m, m-1, \dots, 1$, respectively. Let $p_k$ denote the ranking point assigned to selector $h_k$, and $h_{\mathrm{opt}}$ denote the best selector for that test case. The score of $h_k$ is then defined as
\begin{align*}
    \mathrm{Score}(h_k) = \frac{p_k}{m} \frac{\mathrm{MISE}(h_{\mathrm{opt}})}{\mathrm{MISE}(h_k)}.
\end{align*}
The overall score of each selector is obtained by summing its scores over the considered cases, as shown in Table~\ref{tab:rank}. Each entry reports the total score, with the Dirichlet and Gaussian-ilr scores shown in parentheses and summing to the total. For density estimation $(r=0)$, $\hat{h}_{\mathrm{EMI}}$ is the most competitive selector when $n=500$, while $\hat{h}_{\mathrm{LCV}}$ and $\hat{h}_{\mathrm{RSLS}}$ achieve the highest scores for $d=2$ and $d=4$, respectively, when $n=100$. For first-derivative estimation $(r=1)$, $\hat{h}_{\mathrm{EMI}}$ shows a similar pattern, while $\hat{h}_{\mathrm{RSLS}}$ performs best when $n=100$. Overall, $\hat{h}_{\mathrm{LSCV}}$ performs the worst for both density and first-derivative estimation.

\section{Real data applications}
\label{sec:app}

\subsection{Density estimation for wholesale customers data}
\label{sec:app_wholesale}

The Wholesale Customers dataset \citep{Cardoso2013} comes from the marketing and management domain and is publicly available from the UCI Machine Learning Repository. The dataset contains observations on 440 customers of a wholesale distributor, of whom 298 belong to the Horeca channel, corresponding to hotels, restaurants, and cafés, and the remaining 142 belong to the Retail channel. For each customer, the dataset records annual spending in monetary units on six categories of products: fresh, milk, grocery, frozen, detergents \& paper, and delicatessen. Here, we focus on the relative allocation of expenditure across these categories. Therefore, each observation is transformed into a six-part composition by applying the closure operation.

Since the true density is unknown, the MISE and MIAE cannot be evaluated directly for real data. We therefore compare the selectors using repeated $K$-fold LSCV. For a given partition $I_1, \dots, I_K$, let $n_k = |I_k|$ and $\hat{f}_h^{-I_k}$ denote the estimator fitted using all observations except those in $I_k$. For each fold and each selector, the bandwidth is re-selected using only these $n-n_k$ training observations. The $K$-fold LSCV is
\begin{align*}
    \hat{R}_\mathrm{CV} = \frac{1}{K} \sum_{k=1}^K\Big[\int_{\Delta^d} \hat{f}_h^{-I_k}(\xb)^2 \,\lambda_d(\rd\xb)  - \frac{2}{n_k} \sum_{i \in I_k} \hat{f}_h^{-I_k}(\Xb_i)\Big].
\end{align*}
The integral part is available in closed form because $\hat{f}_h^{-I_k}$ is a finite mixture of Dirichlet densities. Conditional on the training data, the validation average is an unbiased estimate of $\int_{\Delta^d} f(\xb)\hat{f}_h^{-I_k}(\xb)\,\lambda_d(\rd\xb)$. Thus, each fold-specific criterion estimates the data-dependent part of $\mathrm{ISE}[\hat{f}_h^{-I_k}]$ and, for equally sized folds, the expectation of $\hat{R}_\mathrm{CV}$ corresponds to the MISE at sample size $n(1-1/K)$, up to an additive constant common to all estimators. We repeat the $K$-fold procedure $B$ times with independent random partitions and average the resulting values of $\hat{R}_\mathrm{CV}$, which reduces the variability induced by a particular partition.

Table~\ref{tab:wholesale} reports the averaged $\hat{R}_{\mathrm{CV}}$ obtained from 100 repetitions of 10-fold cross-validation. The first row presents the results based on the full dataset ($n=440$), while the second and third rows correspond to the Horeca ($n=298$) and Retail ($n=142$) channels, respectively. $\hat{h}_\mathrm{EMI}$ achieves the best performance for the full dataset and the Horeca channel, whereas the $\hat{h}_\mathrm{LSCV}$ gives the best result for the Retail channel. For the full dataset, the performance of $\hat{h}_\mathrm{EMI}$ and $\hat{h}_\mathrm{LSCV}$ is very close. Compared to the simulation study, $\hat{h}_\mathrm{LSCV}$ is more competitive in this application. One possible reason is that the data contain no observations that are extremely close to one another, making $\hat{h}_\mathrm{LSCV}$ stable enough. The relatively poor performance of $\hat{h}_\mathrm{LCV}$ may partly reflect in loss functions, since $\hat R_{\mathrm{CV}}$ is based on squared-error loss, whereas $\hat{h}_\mathrm{LCV}$ targets Kullback--Leibler loss.

\begin{table}[!h]
\centering
\caption{\small Averaged $\hat{R}_\mathrm{CV}$ with Monte Carlo standard errors (subscripts) over 100 repetitions for the wholesale data, scaled by $10^{-3}$. Boldface indicates the minimum and values not significantly different from it in each scenario.}
\label{tab:wholesale}
\begin{tabular}{lc|cccc}
\toprule
 & $n$ & EMI & LSCV & RSLS & LCV \\
\midrule
Full & 440 & $\mathbf{-7.95}_{0.02}$  & $-7.92_{0.03}$  & $-7.58_{0.02}$  & $-5.81_{0.01}$   \\
Horeca & 298 & $\mathbf{-13.75}_{0.05}$ & $-13.56_{0.06}$ & $-13.06_{0.05}$ & $-8.42_{0.01}$ \\
Retail & 142 & $-9.19_{0.03}$  & $\mathbf{-9.59}_{0.04}$  & $-8.74_{0.03}$  & $-7.99_{0.02}$ \\
\bottomrule
\end{tabular}
\end{table}

\begin{figure}[!h]
    \centering
    \includegraphics[width=0.55\textwidth, trim=2.5cm 0.7cm 1.6cm 0.5cm, clip]{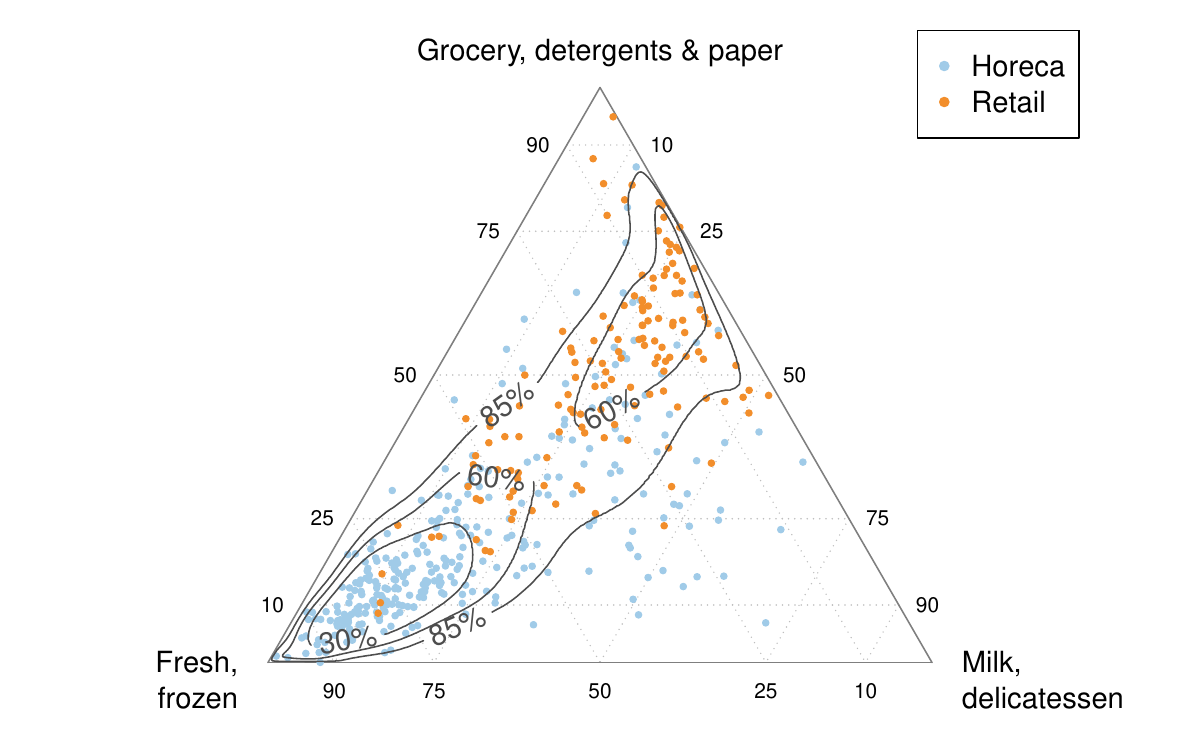}
    \caption{\small Wholesale customer expenditure compositions and estimated density on $\Delta^2$ after amalgamation.}
    \label{fig:wholesale}
\end{figure}

To visualize the data and the estimated density, we amalgamate the six categories into three parts: fresh and frozen, milk and delicatessen, and grocery and detergents \& paper. The estimated density can be mapped exactly onto $\Delta^2$ by the aggregation property \eqref{eq:amalgamation} of the Dirichlet kernel. Figure~\ref{fig:wholesale} shows the amalgamated observations from the full dataset and the resulting EMI density estimate, with observations colored by channel. Although the channel labels were not used in estimation, the two main high-density regions roughly align with the concentrations of Horeca and Retail observations. Horeca customers typically allocate more than $50\%$ of their expenditure to fresh and frozen products and less than $30\%$ to grocery and detergents \& paper. Retail customers show the opposite pattern, with typical shares above $40\%$ for grocery and detergents \& paper and below $30\%$ for fresh and frozen. Differences in milk and delicatessen expenditure are less pronounced.

\subsection{Clustering for sectoral employment data}

The sectoral employment shares from the World Development Indicators of \cite{WorldBankWDI} measure employment in agriculture, industry, and services, each expressed as a percentage of total employment. Since these variables jointly describe the allocation of total employment across the main economic sectors, each economy is represented as a three-part compositional vector. We use the most recent data from 2025. After removing aggregate entries and observations with missing values, the final sample consists of 182 economies.

To identify economies with similar sectoral employment structures, we apply the clustering approach from Section~\ref{sec:cluster}. We do so by estimating the projected first derivative using $\hat{h}_{\mathrm{EMI}}$ based on $K_\mathrm{Atc}$ with $c=\tfrac32$. To reduce spurious small clusters, those containing no more than $5\%$ of the observations are merged into a major cluster. For each point in a small cluster, we find its nearest point in each major cluster and average these distances. The major cluster with the smallest average distance is selected as the merging target.

\begin{figure}[!h]
    \centering
    \includegraphics[page=1, width=0.46\textwidth, trim=1cm 0.9cm 0.8cm 0.7cm, clip]{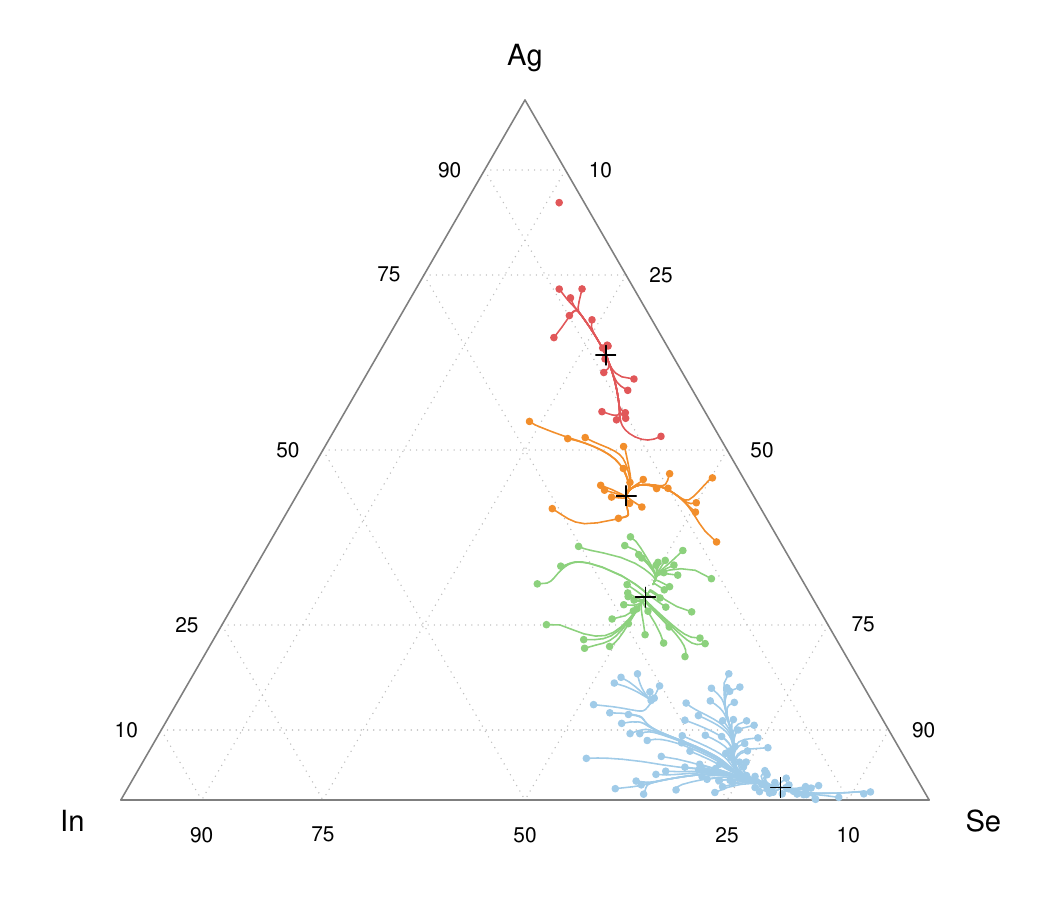}
    \hspace{0.05\textwidth}
    \includegraphics[page=2, width=0.46\textwidth, trim=1cm 0.9cm 0.8cm 0.7cm, clip]{img/Employment.pdf}
    \caption{\small Clustering results for the sectoral employment data. The left panel shows the density-ascent trajectories from the observed employment compositions to their estimated modes, colored by cluster assignment. The right panel shows the estimated basins of attraction, with transparency depending on the density, representative economies labeled by their ISO2C codes, and the economy closest to the cluster mode highlighted in blue.}
    \label{fig:employment}
\end{figure}

Figure~\ref{fig:employment} summarizes the clustering results for the sectoral employment data. The left panel shows the density-ascent trajectories from the observations to their estimated modes. After the post-processing step, the procedure yields four clusters. The trajectories are colored according to their clusters, and the cluster modes are marked by ``$+$'' symbols. The right panel shows the estimated basins of attraction, with each color corresponding to a cluster. Several representative economies are labeled by their ISO2C codes. Among them, the economy closest to its cluster mode is highlighted in blue.

The main differences among the clusters are driven by the relative employment shares along the agriculture--services axis. Moving from the red to the blue cluster, the agricultural employment share generally decreases, whereas the services share increases. Clusters centered at higher services shares also tend to contain a larger number of economies.

\begin{figure}[!h]
    \centering
    \includegraphics[width=0.65\textwidth, trim=0.8cm 3.1cm 0.8cm 3.7cm, clip]{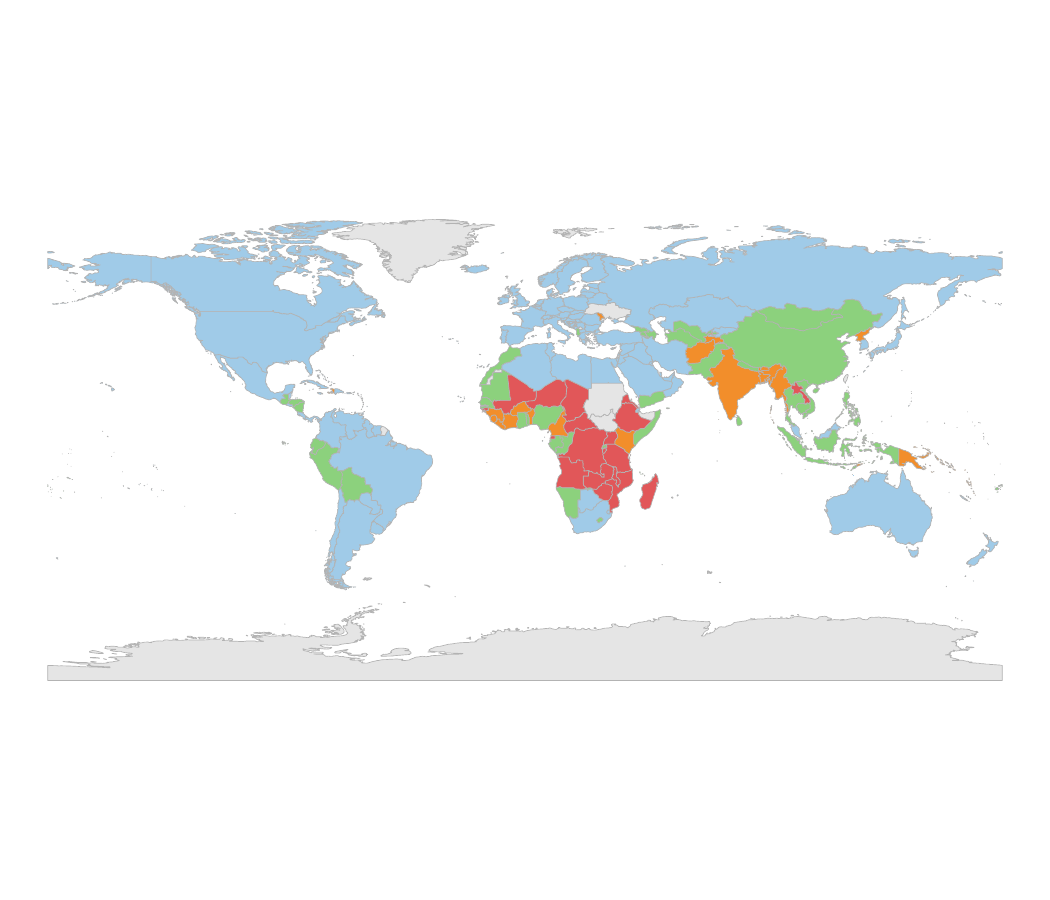}
    \caption{\small The geographical distribution of the cluster assignments on a world map.}
    \label{fig:map}
\end{figure}

The same colors are used in the world map to show the geographical distribution of the cluster assignments, as shown in Figure~\ref{fig:map}. The blue cluster, represented by the United States (US), has the highest service-sector employment share and is concentrated mainly in Europe, the Americas, and Oceania, with some economies also located in Asia, such as Japan (JP), Malaysia (MY), and Iran (IR), and in northern and southern Africa, such as Algeria (DZ) and South Africa (ZA). Most economies in this cluster are high-income. The green and orange clusters, represented by Thailand (TH) and Kenya (KE), respectively, include many developing economies and are widely distributed across Asia and Africa, with a few economies also located in South America, such as Peru (PE). The red cluster, represented by Tanzania (TZ), has the largest agricultural employment share and consists mostly of low-income economies in Africa. Uncolored regions indicate missing data. These patterns suggest that differences in sectoral employment shares are related to both geography and levels of economic development.

\section{Concluding remarks}
\label{sec:conclusion}

We developed bandwidth selection methods for Dirichlet kernel estimation of densities and density derivatives on the simplex. Using a Dirichlet mixture as a reference model, we derived an exact MISE representation and defined the resulting EMI selector. The reference mixture is fitted using a constrained EM algorithm. We also derived an LSCV expression that uses the product formula for Dirichlet densities and avoids explicit numerical integration. To mitigate the variability in LSCV, we introduced the RSLS selector, which replaces the leave-one-out average with a fitted Dirichlet mixture as a smooth reference density. Finally, we derived projected derivatives of the Dirichlet density and their integrability conditions, extended the EMI, LSCV, and RSLS selectors to projected first-derivative estimation, and adapted kernel mean shift clustering to the simplex.

The simulation studies considered Dirichlet targets and Gaussian-ilr targets across dimensions $d=2,4$ and sample sizes $n=100,500$. EMI performed best for Dirichlet targets and was the most competitive selector overall for $n=500$, while RSLS performed best in most settings with $n=100$. RSLS was considerably more stable than LSCV, especially for first-derivative estimation. LCV remained competitive for Gaussian-ilr targets in density estimation. In the real data applications, EMI and LSCV gave the best density estimates for the wholesale customer data, and kernel mean shift clustering with the EMI selector produced interpretable groups of economies based on their sectoral employment shares.

The proposed framework has several limitations. It currently requires strictly positive compositions, so observations with zero components need separate treatment. Under the current Dirichlet kernel parameterization, kernels can differ in peak height and spread depending on their locations. A single global bandwidth therefore gives different amounts of smoothing across the simplex. Finally, for density-derivative estimation, the integrability condition requires $c\geq\tfrac32$. When $c>1$, the mode of $K_{\mathrm{Atc}}$ does not coincide with the observation but is displaced toward the barycenter of the simplex. This shift may introduce location-dependent bias in estimated structural features, such as modes, particularly near the simplex boundary. Future work could explore alternative kernel parameterizations to address these issues while retaining integrability and manageable computation.

\section*{Acknowledgments}

The authors acknowledge financial support from grant PID2024-158399NB-I00 funded by MICIU/AEI/\-10.13039/\-501100011033 and ERDF/EU.

\appendix

\section{Proofs of the main results}
\label{sec:prf_main}

\begin{proof}[Proof of Proposition~\ref{prp:exact_mise}.]
The proof relies mainly on Lemma~\ref{lem:product_dir}, which is applied repeatedly to evaluate the product integrals arising in the three-term decomposition of the MISE:
\begin{align*}
    \mathrm{MISE}_m(h) &= \mathbb{E}\big[\mathrm{ISE}(\hat{f}_h)\big] = \mathbb{E}\left[\int_{\Delta^d}\left(\hat{f}_h(\xb) - f_m(\xb)\right)^2 \,\lambda_d(\rd\xb)\right]\\
    &= \int_{\Delta^d} \mathbb{E}\big[\hat{f}_h(\xb)^2\big]\,\lambda_d(\rd\xb) - 2 \int_{\Delta^d} \mathbb{E}\big[\hat{f}_h(\xb)\big]f_m(\xb)\,\lambda_d(\rd\xb) + \int_{\Delta^d} f_m(\xb)^2 \,\lambda_d(\rd\xb).
\end{align*}

We evaluate each term separately. The third term follows immediately from \eqref{eq:int_2dir}:
\begin{align*}
    \int_{\Delta^d} f_m(\xb)^2 \,\lambda_d(\rd\xb) = \sum_{i,j=1}^m \pi_i \pi_j \int_{\Delta^d} f_\Dir(\xb;\alphab_i) f_\Dir(\xb;\alphab_j)\,\lambda_d(\rd\xb) = \sum_{i,j=1}^m\pi_i \pi_j\frac{C_\Dir(\alphab_i)C_\Dir(\alphab_j)}{C_\Dir(\alphab_i+\alphab_j-\oneb)}.
\end{align*}
For the second term, we first compute the expectation
\begin{align*}
    \mathbb{E}\big[\hat{f}_h(\xb)\big] &= 
    \mathbb{E}\bigg[f_\Dir\Big(\xb; \frac{\Xb_1}{h}+\oneb\Big)\bigg]
    = \sum_{j=1}^m \pi_j\int_{\Delta^d} f_\Dir\Big(\xb; \frac{\yb}{h}+\oneb\Big) f_\Dir(\yb;\alphab_j)\,\lambda_d(\rd\yb).
\end{align*}
Substituting the above result into the second term and applying Fubini's theorem and \eqref{eq:int_2dir}, we obtain
\begin{align*}
    \int_{\Delta^d} \mathbb{E}\big[\hat{f}_h(\xb)\big]f_m(\xb)\,\lambda_d(\rd\xb) &= \int_{\Delta^d} \bigg\{\sum_{j=1}^m \pi_j\int_{\Delta^d} f_\Dir\Big(\xb; \frac{\yb}{h}+\oneb\Big) f_\Dir(\yb;\alphab_j)\,\lambda_d(\rd\yb)\bigg\} \,f_m(\xb)\,\lambda_d(\rd\xb)\\
    &= \sum_{i,j=1}^m \pi_i\pi_j\int_{\Delta^d} \left\{\int_{\Delta^d} f_\Dir\left(\xb; \frac{\yb}{h}+\oneb\right)f_\Dir(\xb;\alphab_i) \,\lambda_d(\rd\xb)\right\} f_\Dir(\yb;\alphab_j) \,\lambda_d(\rd\yb)\\
    &= \sum_{i,j=1}^m \pi_i\pi_j\int_{\Delta^d} \frac{C_{\Dir}(\yb/h+\oneb)C_{\Dir}(\alphab_i)}{C_{\Dir}(\yb/h+\alphab_i)} f_\Dir(\yb;\alphab_j) \,\lambda_d(\rd\yb).
\end{align*}
Applying \eqref{eq:2dir}, the expectation in the first term can be written as
\begin{align*}
    \mathbb{E}\big[\hat{f}_h(\xb)^2\big] 
    &= \frac{1}{n}\,\mathbb{E}\bigg[f_\Dir\Big(\xb; \frac{\Xb}{h}+\oneb\Big)^2\bigg]+\frac{n(n-1)}{n^2}\, \mathbb{E}\bigg[f_\Dir\Big(\xb; \frac{\Xb}{h}+\oneb\Big)\bigg]^2\\
    &= \frac{1}{n} \sum_{j=1}^m \pi_j\int_{\Delta^d} f_\Dir\left(\xb; \frac{\yb}{h}+\oneb\right)^2 f_\Dir(\yb;\alphab_j) \,\lambda_d(\rd\yb)\\
    &~~~~ +\frac{n(n-1)}{n^2} \bigg\{\sum_{j=1}^m \pi_j\int_{\Delta^d}f_\Dir\left(\xb; \frac{\yb}{h}+\oneb\right)f_\Dir(\yb;\alphab_j) \,\lambda_d(\rd\yb)\bigg\}^2\\
    &= \frac{1}{n} \sum_{j=1}^m \pi_j\int_{\Delta^d} 
    \frac{C_{\Dir}(\yb/h+\oneb)^2}{C_{\Dir}(2\yb/h+\oneb)}
    f_\Dir\Big(\xb; \frac{2\yb}{h}+\oneb\Big) f_\Dir(\yb;\alphab_j) \,\lambda_d(\rd\yb)\\
    &~~~~ +(1-n^{-1}) \sum_{i,j=1}^m \pi_i\pi_j\int_{\Delta^d}\int_{\Delta^d}f_\Dir\left(\xb; \frac{\yb_1}{h}+\oneb\right)f_\Dir\left(\xb; \frac{\yb_2}{h}+\oneb\right)\\
    &~~~~ \times f_\Dir(\yb_1;\alphab_i) f_\Dir(\yb_2;\alphab_j) \,\lambda_d(\rd\yb_1)\,\lambda_d(\rd\yb_2).
\end{align*}
Integrating the expectation and applying Fubini's theorem together with \eqref{eq:int_2dir}, we obtain
\begin{align*}
    \int_{\Delta^d} \mathbb{E}\big[\hat{f}_h(\xb)^2\big]\,\lambda_d(\rd\xb) &= \frac{1}{n} \sum_{j=1}^m \pi_j\int_{\Delta^d} 
    \frac{C_{\Dir}(\yb/h+\oneb)^2}{C_{\Dir}(2\yb/h+\oneb)}
    \left\{\int_{\Delta^d} f_\Dir\Big(\xb; \frac{2\yb}{h}+\oneb\Big)\,\lambda_d(\rd\xb)\right\} f_\Dir(\yb;\alphab_j) \,\lambda_d(\rd\yb)\\
    &~~~~ +(1-n^{-1}) \sum_{i,j=1}^m \pi_i\pi_j\int_{\Delta^d}\int_{\Delta^d}\left\{\int_{\Delta^d} f_\Dir\left(\xb; \frac{\yb_1}{h}+\oneb\right)f_\Dir\left(\xb; \frac{\yb_2}{h}+\oneb\right)\,\lambda_d(\rd\xb)\right\}\\
    &~~~~ \times f_\Dir(\yb_1;\alphab_i) f_\Dir(\yb_2;\alphab_j) \,\lambda_d(\rd\yb_1)\,\lambda_d(\rd\yb_2)\\
    &= \frac{1}{n} \sum_{j=1}^m \pi_j\int_{\Delta^d} 
    \frac{C_{\Dir}(\yb/h+\oneb)^2}{C_{\Dir}(2\yb/h+\oneb)}
    f_\Dir(\yb;\alphab_j) \,\lambda_d(\rd\yb)\\
    &~~~~ +(1-n^{-1}) \sum_{i,j=1}^m \pi_i\pi_j\int_{\Delta^d}\int_{\Delta^d}\frac{C_{\Dir}(\yb_1/h+\oneb)C_{\Dir}(\yb_2/h+\oneb)}{C_{\Dir}((\yb_1+\yb_2)/h+\oneb)}\\  
    &~~~~ \times f_\Dir(\yb_1;\alphab_i)f_\Dir(\yb_2;\alphab_j) \,\lambda_d(\rd\yb_1)\,\lambda_d(\rd\yb_2).
\end{align*}

Combining the three terms, we obtain
\begin{align*}
    \mathrm{MISE}_m(h) &= \frac{1}{n} \sum_{j=1}^m \pi_j\int_{\Delta^d} 
    \frac{C_{\Dir}(\yb/h+\oneb)^2}{C_{\Dir}(2\yb/h+\oneb)}
    f_\Dir(\yb;\alphab_j) \,\lambda_d(\rd\yb)\\
    &~~~~ +(1-n^{-1}) \sum_{i,j=1}^m \pi_i\pi_j\int_{\Delta^d}\int_{\Delta^d}\frac{C_{\Dir}(\yb_1/h+\oneb)C_{\Dir}(\yb_2/h+\oneb)}{C_{\Dir}((\yb_1+\yb_2)/h+\oneb)} \\  
    &~~~~ \times f_\Dir(\yb_1;\alphab_i)f_\Dir(\yb_2;\alphab_j) \,\lambda_d(\rd\yb_1)\,\lambda_d(\rd\yb_2) \\
    &~~~~ -2\sum_{i,j=1}^m \pi_i\pi_j\int_{\Delta^d} \frac{C_{\Dir}(\yb/h+\oneb)C_{\Dir}(\alphab_i)}{C_{\Dir}(\yb/h+\alphab_i)} f_\Dir(\yb;\alphab_j) \,\lambda_d(\rd\yb)\\
    &~~~~ + \sum_{i,j=1}^m\pi_i \pi_j\frac{C_\Dir(\alphab_i)C_\Dir(\alphab_j)}{C_\Dir(\alphab_i+\alphab_j-\oneb)}\\
    &= n^{-1} C_d(h)+\boldsymbol{\pi}^\top\big[(1-n^{-1})\boldsymbol{\Psi}_2(h)-2\boldsymbol{\Psi}_1(h)+\boldsymbol{\Psi}_0(h)\big]\boldsymbol{\pi}.
\end{align*}
\end{proof}

\begin{proof}[Proof of Proposition~\ref{prp:lscv}.]
    For the first term of LSCV, by \eqref{eq:int_2dir}, we have
    \begin{align*}
    \int_{\Delta^d} \hat{f}_h(\xb)^2\,\lambda_d(\rd\xb) &= \frac{1}{n^2}\sum_{i,j=1}^n \int_{\Delta^d} f_\Dir\Big(\xb; \frac{\Xb_i}{h}+\oneb\Big)f_\Dir\Big(\xb; \frac{\Xb_j}{h}+\oneb\Big)\,\lambda_d(\rd\xb)\\
    &=\frac{1}{n^2}\sum_{i,j=1}^n \frac{C_{\Dir}(\Xb_i/h+\oneb)C_{\Dir}(\Xb_j/h+\oneb)}{C_{\Dir}((\Xb_i+\Xb_j)/h+\oneb)}\\
    &=\frac{1}{n^2}\frac{\Gamma(1/h+D)^2}{\Gamma(2/h+D)}\sum_{i,j=1}^n \prod_{k=1}^D \frac{\Gamma((X_{ik}+X_{jk})/h+1)}{\Gamma(X_{ik}/h+1)\Gamma(X_{jk}/h+1)}.
\end{align*}
The second term can be written as
\begin{align*}
    \frac{1}{n}\sum_{i=1}^n \hat{f}^{-i}_h(\Xb_i)&=\frac{1}{n(n-1)}\sum_{i=1}^n\sum_{\substack{j=1\\j\neq i}}^n f_{\Dir}\Big(\Xb_i;\frac{\Xb_j}{h}+\oneb\Big)\\
    &=\frac{1}{n(n-1)}\sum_{\substack{i,j=1\\i\neq j}}^n \frac{\Gamma(1/h+D)}{\prod_{k=1}^D\Gamma(X_{jk}/h+1)}\prod_{k=1}^D X_{ik}^{X_{jk}/h}\\
    &=\frac{\Gamma(1/h+D)}{n(n-1)}\sum_{i,j=1}^n (1-\delta_{ij})\prod_{k=1}^D \frac{X_{ik}^{X_{jk}/h}}{\Gamma(X_{jk}/h+1)}.
\end{align*}
Combining the two terms, we obtain
\begin{align*}
    \mathrm{LSCV}(h)&=\frac{\Gamma(1/h+D)}{n^2}\sum_{i,j=1}^n \bigg\{
    \frac{\Gamma(1/h+D)}{\Gamma(2/h+D)} \prod_{k=1}^D \frac{\Gamma((X_{ik}+X_{jk})/h+1)}{\Gamma(X_{ik}/h+1)\Gamma(X_{jk}/h+1)}\\
    &~~~~ -2(1-n^{-1})^{-1}(1-\delta_{ij})\prod_{k=1}^D \frac{X_{ik}^{X_{jk}/h}}{\Gamma(X_{jk}/h+1)}\bigg\}.
\end{align*}
\end{proof}

\begin{proof}[Proof of Corollary~\ref{coro:rsls}.]
The first term is the same as that in the LSCV and has been derived in the proof of Proposition~\ref{prp:lscv}. For the second term, by \eqref{eq:int_2dir}, we have
\begin{align*}
    \int_{\Delta^d} \hat{f}_h(\xb)f_m(\xb) \,\lambda_d(\rd\xb) &= \frac{1}{n}\sum_{i=1}^n \sum_{j=1}^m \pi_j \int_{\Delta^d} f_\Dir\Big(\xb; \frac{\Xb_i}{h}+\oneb\Big) f_\Dir(\xb; \alphab_j) \,\lambda_d(\rd\xb)\\
    &= \frac{1}{n}\sum_{i=1}^n \sum_{j=1}^m \pi_j \frac{C_\Dir(\Xb_i/h+\oneb)C_\Dir(\alphab_j)}{C_\Dir(\Xb_i/h+\alphab_j)}.
\end{align*}
\end{proof}

\begin{proof}[Proof of Proposition~\ref{prp:integrability}.]
Obviously, $\Dsf^{\otimes r}f_\Dir \in L^p(\Delta^d)$ if and only if each component belongs to $L^p(\Delta^d)$. For any $r \geq 0$, each component of $\mathsf{D}^{\otimes r} f_\Dir(\xb; \alphab)$ is
\begin{align*}
   C_{\boldsymbol{s}}(\alphab)\prod_{k=1}^D x_k^{\alpha_k - s_k - 1}, ~~~~ \boldsymbol{s} \in \mathbb{N}_0^D,~ \boldsymbol{s}^\top\oneb = r,
\end{align*}
where $C_{\boldsymbol{s}}(\alphab)$ is a constant. 

We first prove part~\ref{prop:integrability-i}. If $\alpha_k > r$ for all $k$, then $\alpha_k - s_k \ge \alpha_k - r > 0$. Each component is integrable on $\Delta^d$ with
\begin{align*}
    \int_{\Delta^d}
    \Bigl| C_{\boldsymbol s}(\alphab)
    \prod_{k=1}^D x_k^{\alpha_k-s_k-1} \Bigr|\,\lambda_d(\rd\xb)
    = \left|C_{\boldsymbol s}(\alphab)\right|
    C_\Dir(\alphab-\boldsymbol s)^{-1}
    = \left|C_{\boldsymbol s}(\alphab)\right|
    \frac{\prod_{k=1}^D\Gamma(\alpha_k-s_k)}
    {\Gamma(\alpha_0-r)} <\infty.
\end{align*}
Hence, $\mathsf{D}^{\otimes r} f_\Dir(\xb; \alphab) \in L^1(\Delta^d)$. We now prove part~\ref{prop:integrability-ii}. Squaring the generic component above gives
\begin{align*}
    C_{\boldsymbol{s}}(\alphab)^2 \prod_{k=1}^D x_k^{2(\alpha_k - s_k - 1)} = C_{\boldsymbol{s}}(\alphab)^2
  \prod_{k=1}^D x_k^{\beta_k - 1},
\end{align*}
where $\beta_k := 2\alpha_k - 2s_k - 1$. If $\alpha_k > r + \tfrac12$ for all $k$, then $s_k\le r$ implies $\beta_k>0$. So each squared component is integrable on $\Delta^d$, and hence $\mathsf{D}^{\otimes r} f_\Dir\in L^2(\Delta^d)$. 
\end{proof}

\begin{proof}[Proof of Corollary~\ref{coro:exact_mise_r1}.]
The first-derivative MISE admits the following three-term decomposition:
\begin{align*}
    \mathrm{MISE}_{1,m}(h) = \int_{\Delta^d} \mathbb{E}\big[\big\|\tDsf\hat{f}_h(\xb)\big\|^2\big]\,\lambda_d(\rd\xb) - 2 \int_{\Delta^d} \mathbb{E}\big[\tDsf\hat{f}_h(\xb)\big]^\top\,\tDsf f_m(\xb)\,\lambda_d(\rd\xb) + \int_{\Delta^d} \big\|\tDsf f_m(\xb)\big\|^2 \,\lambda_d(\rd\xb).
\end{align*}
The remaining proof steps follow those in the proof of Proposition~\ref{prp:exact_mise}, using Lemma~\ref{lem:product_dir_r1} in place of Lemma~\ref{lem:product_dir}. Applying Fubini's theorem and Lemma~\ref{lem:product_dir_r1} to these three terms, and then collecting the resulting expressions, we obtain $\mathrm{MISE}_{1,m}(h) = n^{-1} C_d(h)+\boldsymbol{\pi}^\top\big[(1-n^{-1})\boldsymbol{\Psi}_2(h)-2\boldsymbol{\Psi}_1(h)+\boldsymbol{\Psi}_0(h)\big]\boldsymbol{\pi}$.
\end{proof}

\begin{proof}[Proof of Proposition~\ref{prp:int_by_parts}.]
We rewrite \eqref{eq:int_by_parts} in the coordinates $\zb = (x_1, \ldots, x_d)$ of $\lambda_d$, where it becomes a standard integration by parts on a domain of $\Rb^d$ whose boundary terms vanish by \ref{prp:int_by_parts_2}.

With $\Omega$ and $\iota$ as in Section~\ref{sec:Dir_KDE}, let $F := f \circ \iota$ and $G := g \circ \iota$, and let $\Dsf F$ and $\Hess G$ be the gradient (column vector) and the Hessian with respect to $\zb \in \Omega$. Since $\iota$ is affine with Jacobian $\Jb := (\eb_1 - \eb_D, \ldots, \eb_d - \eb_D)$, the chain rule gives $\Dsf F = \Jb^\top \Dsf f$ and $\Hess G = \Jb^\top \Hess g \, \Jb$. Since the columns of $\Jb$ span the tangent space $\{\vb \in \Rb^D : \oneb^\top \vb = 0\}$, $\Pb$ is the hat matrix of $\Jb$: $\Pb = \Jb \Wb \Jb^\top$ with $\Wb := (\Jb^\top \Jb)^{-1}$. Hence
\begin{align*}
  \tDsf f^\top \tDsf g &= \Dsf f^\top \Pb \, \Dsf g = \Dsf F^\top \Wb \, \Dsf G, \\
  \vecop(\Pb)^\top \Dsf^{\otimes 2} g &= \tr(\Pb \Hess g) = \tr(\Wb \, \Hess G),
\end{align*}
and, by \eqref{eq:measure} and the product rule, \eqref{eq:int_by_parts} is equivalent to
\begin{align}
  \sum_{i,j=1}^d W_{ij} \int_\Omega \big(\partial_i F \, \partial_j G + F \, \partial_i \partial_j G\big) \,\rd\zb = \sum_{i,j=1}^d W_{ij} \int_\Omega \partial_i (F \, \partial_j G) \,\rd\zb = 0, \label{eq:ibp_ij}
\end{align}
where $\partial_i := \partial/\partial z_i$. All these integrands are integrable by \ref{prp:int_by_parts_1} and the Cauchy--Schwarz inequality, since $\Pb\Jb = \Jb$ gives $\Dsf F = \Jb^\top \tDsf f$, $\Dsf G = \Jb^\top \tDsf g$, and $\vecop(\Hess G) = (\Jb \otimes \Jb)^\top \tDsf^{\otimes 2} g$.

To prove \eqref{eq:ibp_ij}, we show that $\int_\Omega \partial_i (F \, \partial_j G) \,\rd\zb = 0$ for all $i, j$. We integrate first in $z_i$, which runs over $(0, s)$ with $s := 1 - \sum_{m \neq i} z_m$. We have that
\begin{align*}
  \int_0^s \partial_i (F \, \partial_j G) \,\rd z_i = \lim_{z_i \uparrow s} F \, \partial_j G - \lim_{z_i \downarrow 0} F \, \partial_j G = 0.
\end{align*}
Indeed, $\iota(\zb)$ tends to a point of $\partial\Delta^d$ in both limits, where $x_D \to 0$ and $x_i \to 0$, respectively, and $F \, \partial_j G = (\eb_j - \eb_D)^\top f \, \tDsf g$ at $\xb = \iota(\zb)$ tends to zero there by \ref{prp:int_by_parts_2}. Integrating over the remaining coordinates gives the claim.

For the last claim, by linearity it suffices to check \ref{prp:int_by_parts_1} and \ref{prp:int_by_parts_2} for $g = f_\Dir(\cdot; \alphab)$ with $\alphab = \Xb_i / h + c\oneb > 2\oneb$, which is smooth on $(0, \infty)^D$. By \eqref{eq:Dir_gradient}, the $j$th entry of $\Dsf g$ is a constant times $x_j^{\alpha_j - 2} \prod_{m \neq j} x_m^{\alpha_m - 1}$, whose exponents are positive. Hence $\tDsf g = \Pb \Dsf g$ is bounded on $\Delta^d$ and tends to $\mathbf{0}$ at $\partial\Delta^d$, where some $x_m$ vanishes. Also, $\|\tDsf^{\otimes 2} g\| \leq \|\Dsf^{\otimes 2} g\|$ since $\Pb^{\otimes 2}$ is an orthogonal projection, and $\|\Dsf^{\otimes 2} g\|$ is integrable by Proposition~\ref{prp:integrability}\ref{prop:integrability-i} with $r = 2$. As $f$ is bounded and $\|\tDsf f\| \in L^1(\Delta^d)$, \ref{prp:int_by_parts_1} and \ref{prp:int_by_parts_2} follow.
\end{proof}

\begin{proof}[Proof of Corollary~\ref{coro:lscv_r1}.]
As in the proof of Proposition~\ref{prp:lscv}, the first term can be simplified using \eqref{eq:int_2dir_r1}. For the second term, we have
\begin{align*}
    &\frac{1}{n}\vecop(\Pb)^\top
    \sum_{i=1}^n \Dsf^{\otimes 2}\hat{f}^{-i}_h(\Xb_i)
    = \frac{1}{n(n-1)}
    \vecop(\Pb)^\top
    \sum_{i=1}^n\sum_{\substack{j=1\\j\neq i}}^n
    \Dsf^{\otimes 2}
    f_{\Dir}\Big(\Xb_i;\frac{\Xb_j}{h}+c\oneb\Big).
\end{align*}
For $i\neq j$, let $\gb_{ij}=\gb(\Xb_i;\Xb_j/h+c\oneb)$, and recall $\boldsymbol{M}_{ij}$ as defined in Corollary~\ref{coro:lscv_r1}, set
\begin{align*}
    \omega_{ij}(h)
    := \prod_{k=1}^D
    \frac{X_{ik}^{X_{jk}/h+c-1}}{\Gamma(X_{jk}/h+c)}.
\end{align*}
The second derivative can be written as
\begin{align*}
    \Dsf^{\otimes 2}
    f_{\Dir}\Big(\Xb_i;\frac{\Xb_j}{h}+c\oneb\Big)
    &= \Big(\gb_{ij}^{\otimes 2} - \sum_{k=1}^D \frac{X_{jk}/h+c-1}{X_{ik}^2}\,\eb_k^{\otimes 2} \Big)
    f_{\Dir}\Big(\Xb_i;\frac{\Xb_j}{h}+c\oneb\Big)\\
    &= \Gamma(1/h+Dc)\, 
    \vecop(\boldsymbol{M}_{ij})\,
    \omega_{ij}(h).
\end{align*}
Substituting this identity into the second term gives
\begin{align*}
    \frac{\Gamma(1/h+Dc)}{n(n-1)}
    &\sum_{\substack{i,j=1\\i\neq j}}^n
    \vecop(\Pb)^\top
    \vecop(\boldsymbol{M}_{ij})\,
    \omega_{ij}(h)\\
    =\,& \frac{\Gamma(1/h+Dc)}{n(n-1)}
    \sum_{\substack{i,j=1\\i\neq j}}^n
    \Big(\sum_{s,t=1}^D P_{st}M_{ij,st}\Big)
    \,\omega_{ij}(h)\\
    =\,& \frac{\Gamma(1/h+Dc)}{n(n-1)}
    \sum_{i,j=1}^n (1-\delta_{ij})
    \Big(\sum_{s,t=1}^D
    (\delta_{st}-D^{-1})M_{ij,st}\Big)
    \,\omega_{ij}(h).
\end{align*}
Combining these two terms, we obtain
\begin{align*}
    \mathrm{LSCV}_1(h) 
    &= \frac{\Gamma(1/h+Dc)}{n^2}\sum_{i,j=1}^n \bigg\{
    \frac{\Gamma(1/h+Dc)}{\Gamma(2/h+2Dc-
    D)}\,H_{ij}\, \prod_{k=1}^D \frac{\Gamma((X_{ik}+X_{jk})/h+2c-1)}{\Gamma(X_{ik}/h+c)\Gamma(X_{jk}/h+c)}\\
    &~~~~ +2(1-n^{-1})^{-1}(1-\delta_{ij})\Big(\sum_{s,t=1}^D(\delta_{st}-D^{-1}) M_{ij,st}\Big) \Big(\prod_{k=1}^D \frac{X_{ik}^{X_{jk}/h+c-1}}{\Gamma(X_{jk}/h+c)}\Big)\bigg\}.
\end{align*}
\end{proof}

\begin{proof}[Proof of Corollary~\ref{coro:rsls_r1}.]
The result follows similarly to Corollary~\ref{coro:rsls}: the first term is the same as that in $\mathrm{LSCV}_1$, while the second term is obtained by applying \eqref{eq:int_2dir_r1}.
\end{proof}

\section{Lemmata}
\label{sec:prf_lem}

This section collects several technical lemmas that are used in the proofs of the main results. Lemma~\ref{lem:product_dir} gives the product formula for two Dirichlet densities together with the associated integral, whereas Lemma~\ref{lem:product_dir_r1} provides the corresponding result for the projected first derivatives. The former is used in the proofs of Propositions \ref{prp:exact_mise}--\ref{prp:lscv} and Corollary \ref{coro:rsls}, while the latter is used in the proofs of Corollaries \ref{coro:exact_mise_r1}, \ref{coro:lscv_r1}, and \ref{coro:rsls_r1}.

\begin{lemma} \label{lem:product_dir}
Let $\alphab_1,\alphab_2 \in \Rb_+^D$ be such that $\alphab_1+\alphab_2 > \oneb$. Then, for $\xb \in \Delta^d$,
\begin{align}
    f_\Dir(\xb;\alphab_1)\,f_\Dir(\xb;\alphab_2) = \frac{C_\Dir(\alphab_1)C_\Dir(\alphab_2)}{C_\Dir(\alphab_1 + \alphab_2 - \oneb)} f_\Dir(\xb; \alphab_1 + \alphab_2 - \oneb).
    \label{eq:2dir}
\end{align}
    Consequently,
\begin{align}
    \int_{\Delta^d} f_\Dir(\xb;\alphab_1)\,f_\Dir(\xb;\alphab_2)\,\lambda_d(\rd\xb) = \frac{C_\Dir(\alphab_1)C_\Dir(\alphab_2)}{C_\Dir(\alphab_1 + \alphab_2 - \oneb)}.
    \label{eq:int_2dir}
\end{align}
\end{lemma}

\begin{lemma} \label{lem:product_dir_r1}
Let $\alphab_1,\alphab_2 > \oneb$ be such that
$\alphab_1+\alphab_2 > 3\oneb$, and consider the notation introduced in \eqref{eq:Heq}. Then, for $\xb \in \Delta^d$,
\begin{align}
    \tDsf f_\Dir(\xb;\alphab_1)^\top & \tDsf f_\Dir(\xb;\alphab_2) \nonumber\\
    =\,& (\alpha_{0,1}+\alpha_{0,2}-D-1)^{\underline{2}}\,\frac{C_\Dir(\alphab_1)C_\Dir(\alphab_2)}{C_\Dir(\alphab_1+\alphab_2-\oneb)} \nonumber\\
    & \times \Big\{\frac{D-1}{D}\sum_{k=1}^D \frac{(\alpha_{1k}-1)(\alpha_{2k}-1)}{(\alpha_{1k}+\alpha_{2k}-2)^{\underline{2}}} \,f_\Dir(\xb;\alphab_1+\alphab_2-2\eb_k-\oneb) \nonumber\\
    &~~~~ + \frac{1}{D} \sum_{k=1}^D \frac{(\alpha_{1k}-1)(\alpha_{2k}-1)}{(\alpha_{1k}+\alpha_{2k}-2)^2} f_\Dir(\xb;\alphab_1+\alphab_2-2\eb_k-\oneb) \nonumber\\
    &~~~~ -\frac{1}{D} \sum_{k,l=1}^D \frac{(\alpha_{1k}-1)(\alpha_{2l}-1)}{(\alpha_{1k}+\alpha_{2k}-2)(\alpha_{1l}+\alpha_{2l}-2)} f_\Dir(\xb;\alphab_1+\alphab_2-\eb_k-\eb_l-\oneb) \Big\}.
\end{align}
   Consequently,
\begin{align}
    \int_{\Delta^d} \tDsf f_\Dir(\xb;\alphab_1)^\top\tDsf  f_\Dir(\xb;\alphab_2)\,\lambda_d(\rd\xb) = H(\alphab_1, \alphab_2)\,\frac{C_\Dir(\alphab_1)C_\Dir(\alphab_2)}{C_\Dir(\alphab_1+\alphab_2-\oneb)}. \label{eq:int_2dir_r1}
\end{align}
\end{lemma}

\begin{proof}[Proof of Lemma~\ref{lem:product_dir}]
The product of two Dirichlet densities is
\begin{align*}
    f_\Dir(\xb;\alphab_1)\,f_\Dir(\xb;\alphab_2) 
    &= C_\Dir(\alphab_1)C_\Dir(\alphab_2) \prod_{k = 1}^D x_k^{\alpha_{1k} + \alpha_{2k} - 1 - 1}\\
    &= \frac{C_\Dir(\alphab_1)C_\Dir(\alphab_2)}{C_\Dir(\alphab_1 + \alphab_2 - \oneb)} f_\Dir(\xb; \alphab_1 + \alphab_2 - \oneb).
\end{align*}
\end{proof}

\begin{proof}[Proof of Lemma~\ref{lem:product_dir_r1}]
For the $j$th component of $\Dsf f_\Dir(\xb; \alphab)$, we have
\begin{align*}
    \frac{\partial}{\partial x_j} f_\Dir(\xb; \alphab) &= C_\Dir(\alphab)(\alpha_j - 1)\,x_j^{\alpha_j-2} \prod_{k\neq j} x_k^{\alpha_k-1} \\
    &= (\alpha_j - 1)\frac{C_\Dir(\alphab)}{C_\Dir(\alphab-\eb_j)} f_\Dir(\xb; \alphab-\eb_j) \\
    &= (\alpha_0 - 1)\,f_\Dir(\xb; \alphab-\eb_j).
\end{align*}
Hence,
\begin{align}
    \Dsf f_\Dir(\xb;\alphab) = (\alpha_0-1)\big(f_\Dir(\xb; \alphab-\eb_k)\big)_{k=1}^D. \label{eq:Dir_r1}
\end{align}
The product of two projected first derivatives of the Dirichlet density is
\begin{align*}
    \tDsf f_\Dir(\xb;\alphab_1)^\top & \tDsf
    f_\Dir(\xb;\alphab_2)\\
    =\,& \Dsf f_\Dir(\xb;\alphab_1)^\top
    \bigl(\Ib_D-D^{-1}\oneb\oneb^\top\bigr)
    \Dsf f_\Dir(\xb;\alphab_2)\\
    =\,& \Dsf f_\Dir(\xb;\alphab_1)^\top
    \Dsf f_\Dir(\xb;\alphab_2)
    -D^{-1} \bigl(\oneb^\top\Dsf f_\Dir(\xb;\alphab_1)\bigr)
    \bigl(\oneb^\top\Dsf f_\Dir(\xb;\alphab_2)\bigr).
\end{align*}

For the first term, by \eqref{eq:Dir_r1} and \eqref{eq:2dir}, we obtain
\begin{align*}
    \Dsf f_\Dir(\xb;\alphab_1)^\top & \Dsf f_\Dir(\xb;\alphab_2)\\ 
    =\,& (\alpha_{0,1} - 1)(\alpha_{0,2} - 1) \sum_{k=1}^D f_\Dir(\xb; \alphab_1 - \eb_k)f_\Dir(\xb; \alphab_2 - \eb_k)\\
    =\,& (\alpha_{0,1} - 1)(\alpha_{0,2} - 1) \sum_{k=1}^D \frac{C_\Dir(\alphab_1 - \eb_k)C_\Dir(\alphab_2 - \eb_k)}{C_\Dir(\alphab_1 + \alphab_2 - 2\eb_k - \oneb)} f_\Dir(\xb; \alphab_1 + \alphab_2 - 2\eb_k - \oneb)\\
    =\,& (\alpha_{0,1}+\alpha_{0,2}-D-1)^{\underline{2}}\frac{C_\Dir(\alphab_1)C_\Dir(\alphab_2)}{C_\Dir(\alphab_1+\alphab_2-\oneb)} \sum_{k=1}^D \frac{(\alpha_{1k}-1)(\alpha_{2k}-1)}{(\alpha_{1k} + \alpha_{2k}-2)^{\underline{2}}}\\
    &\, \times f_\Dir(\xb;\alphab_1+\alphab_2-2\eb_k-\oneb).
\end{align*}
Similarly, by \eqref{eq:Dir_r1} and \eqref{eq:2dir}, the second term can be rewritten as
\begin{align*}
    \left(\oneb^\top\Dsf f_\Dir(\xb;\alphab_1)\right)&\left(\oneb^\top\Dsf f_\Dir(\xb;\alphab_2)\right)\\ 
    =\,& (\alpha_{0,1} - 1)(\alpha_{0,2} - 1) \sum_{k,l=1}^D f_\Dir(\xb;\alphab_1-\eb_k) f_\Dir(\xb;\alphab_2-\eb_l)\\
    =\,& (\alpha_{0,1} - 1)(\alpha_{0,2} - 1) \sum_{k,l=1}^D \frac{C_\Dir(\alphab_1-\eb_k)C_\Dir(\alphab_2-\eb_l)}{C_\Dir(\alphab_1+\alphab_2-\eb_k-\eb_l-\oneb)}\\ 
    &\, \times f_\Dir(\xb;\alphab_1+\alphab_2-\eb_k-\eb_l-\oneb)\\
    =\,& (\alpha_{0,1}+\alpha_{0,2}-D-1)^{\underline{2}}\,\frac{C_\Dir(\alphab_1)C_\Dir(\alphab_2)}{C_\Dir(\alphab_1+\alphab_2-\oneb)} \\
    & \times\bigg\{\sum_{k=1}^D \frac{(\alpha_{1k}-1)(\alpha_{2k}-1)}{(\alpha_{1k} + \alpha_{2k}-2)^{\underline{2}}}\,f_\Dir(\xb;\alphab_1+\alphab_2-2\eb_k-\oneb) \\
    &~~~~ + \sum_{\substack{k,l=1\\ k\neq l}}^D \frac{(\alpha_{1k}-1)(\alpha_{2l}-1)}{(\alpha_{1k} + \alpha_{2k}-2)(\alpha_{1l} + \alpha_{2l}-2)} \, f_\Dir(\xb;\alphab_1+\alphab_2-\eb_k-\eb_l-\oneb)\bigg\}.
\end{align*}
Combining the two terms gives
\begin{align*}
    \tDsf f_\Dir(\xb;\alphab_1)^\top & \tDsf f_\Dir(\xb;\alphab_2) \\
    =\,& (\alpha_{0,1}+\alpha_{0,2}-D-1)^{\underline{2}}\,\frac{C_\Dir(\alphab_1)C_\Dir(\alphab_2)}{C_\Dir(\alphab_1+\alphab_2-\oneb)}\\
    & \times \bigg\{\frac{D-1}{D}\sum_{k=1}^D \frac{(\alpha_{1k}-1)(\alpha_{2k}-1)}{(\alpha_{1k}+\alpha_{2k}-2)^{\underline{2}}} \,f_\Dir(\xb;\alphab_1+\alphab_2-2\eb_k-\oneb)\\
    &~~~~ -\frac{1}{D} \sum_{k,l=1}^D \frac{(\alpha_{1k}-1)(\alpha_{2l}-1)}{(\alpha_{1k}+\alpha_{2k}-2)(\alpha_{1l}+\alpha_{2l}-2)} f_\Dir(\xb;\alphab_1+\alphab_2-\eb_k-\eb_l-\oneb)\\
    &~~~~ + \frac{1}{D} \sum_{k=1}^D \frac{(\alpha_{1k}-1)(\alpha_{2k}-1)}{(\alpha_{1k}+\alpha_{2k}-2)^2} f_\Dir(\xb;\alphab_1+\alphab_2-2\eb_k-\oneb)
    \bigg\}.
\end{align*}
\end{proof}

\section{Simulation results for MIAE} 
\label{sec:simu_MIAE}

The integrated absolute error (IAE) measures the $L^1$ distance between the estimator and the target density. Its expectation, referred to as the MIAE, is defined for density and first-derivative estimation, respectively, as
\begin{align*}
    \mathrm{MIAE}(h) := \mathbb{E}\left[\int_{\Delta^d}\left|\hat{f}_h(\xb) - f(\xb)\right| \,\lambda_d(\rd\xb)\right],~~~~
    \mathrm{MIAE}_1(h) :=   \mathbb{E}\left[\int_{\Delta^d}\Big\|\widehat{\tDsf f}_h(\xb) - \tDsf f(\xb)\Big\|_1 \,\lambda_d(\rd\xb)\right],
\end{align*}
where $\|\vb\|_1 = \sum_{j=1}^D |v_j|$. Their Monte Carlo approximations are obtained analogously to \eqref{eq:MCmise} and the reference bandwidth $h_{\mathrm{MIAE}}$ is obtained in the same way as $h_{\mathrm{MISE}}$. Tables \ref{tab:MIAE_r0} and \ref{tab:MIAE_r1} report the corresponding results for density and first-derivative estimation, respectively. The results broadly agree with those based on MISE, except for first-derivative estimation with Gaussian-mixture targets at $d=4$ and $n=500$, where $\hat{h}_{\mathrm{EMI}}$ performs best under MIAE, whereas $\hat{h}_{\mathrm{LSCV}}$ performs best under MISE.

\begin{sidewaystable}[p]
    \centering
    \caption{\small MIAE values for the density estimation simulations, with Monte Carlo standard errors shown as subscripts. Boldface indicates the minimum and values not significantly different from it among the data-driven selectors in each scenario. The oracle benchmark is the MIAE selector.}
    \label{tab:MIAE_r0}
    
    \begin{tabular}{llcccccccccccc}
    \toprule
    Setting & Selector & \multicolumn{6}{c}{Dirichlet} & \multicolumn{6}{c}{Gaussian-ilr} \\
    \cmidrule(lr){3-8} \cmidrule(lr){9-14}
    & & (1) & (2) & (3) & (4) & (5) & (6) & (7) & (8) & (9) & (10) & (11) & (12) \\
    \midrule
    
    \multirow{5}{*}{\centering\makecell[l]{$d=2$, \\ $n=100$ \\ [3pt] ($\times 0.1$)}}
    
    & MIAE & $2.07_{0.01}$ & $2.61_{0.01}$ & $2.25_{0.01}$ & $2.93_{0.01}$ & $3.35_{0.01}$ & $2.47_{0.01}$ & $2.35_{0.01}$ & $2.30_{0.01}$ & $2.47_{0.01}$ & $2.67_{0.01}$ & $3.06_{0.01}$ & $3.87_{0.01}$\\
    & EMI & $2.12_{0.01}$ & $\mathbf{2.66}_{0.01}$ & $\mathbf{2.31}_{0.01}$ & $\mathbf{3.03}_{0.01}$ & $\mathbf{3.46}_{0.01}$ & $2.66_{0.02}$ & $\mathbf{2.42}_{0.01}$ & $2.60_{0.02}$ & $2.82_{0.01}$ & $3.16_{0.02}$ & $3.13_{0.01}$ & $4.07_{0.02}$ \\
    & LSCV & $2.33_{0.02}$ & $2.83_{0.02}$ & $2.44_{0.02}$ & $3.12_{0.02}$ & $3.52_{0.02}$ & $2.67_{0.01}$ & $2.50_{0.01}$ & $2.42_{0.02}$ & $2.56_{0.01}$ & $2.80_{0.02}$ & $3.18_{0.01}$ & $4.00_{0.01}$ \\
    & RSLS & $\mathbf{2.11}_{0.01}$ & $2.67_{0.01}$ & $2.32_{0.01}$ & $3.07_{0.01}$ & $3.53_{0.01}$ & $\mathbf{2.59}_{0.01}$ & $2.45_{0.01}$ & $\mathbf{2.36}_{0.01}$ & $\mathbf{2.52}_{0.01}$ & $2.79_{0.01}$ & $3.17_{0.01}$ & $4.13_{0.02}$ \\
    & LCV & $2.20_{0.02}$ & $2.72_{0.02}$ & $2.36_{0.01}$ & $3.07_{0.02}$ & $\mathbf{3.47}_{0.01}$ & $\mathbf{2.60}_{0.01}$ & $2.45_{0.01}$ & $2.41_{0.01}$ & $2.57_{0.01}$ & $\mathbf{2.75}_{0.01}$ & $\mathbf{3.11}_{0.01}$ & $\mathbf{3.96}_{0.01}$ \\

    \midrule

    \multirow{5}{*}{\centering\makecell[l]{$d=2$, \\ $n=500$ \\ [3pt] ($\times 0.1$)}}
    
    & MIAE & $1.28_{0.01}$ & $1.60_{0.01}$ & $1.36_{0.01}$ & $1.79_{0.01}$ & $2.05_{0.01}$ & $1.54_{0.01}$ & $1.47_{0.01}$ & $1.44_{0.01}$ & $1.58_{0.01}$ & $1.67_{0.01}$ & $1.98_{0.01}$ & $2.37_{0.01}$\\
    & EMI & $1.30_{0.01}$ & $1.61_{0.01}$ & $\mathbf{1.38}_{0.01}$ & $\mathbf{1.81}_{0.01}$ & $\mathbf{2.08}_{0.01}$ & $\mathbf{1.56}_{0.01}$ & $\mathbf{1.49}_{0.01}$ & $\mathbf{1.44}_{0.01}$ & $1.83_{0.01}$ & $\mathbf{1.69}_{0.01}$ & $2.00_{0.01}$ & $\mathbf{2.39}_{0.01}$ \\
    & LSCV & $1.35_{0.01}$ & $1.66_{0.01}$ & $1.42_{0.01}$ & $1.85_{0.01}$ & $2.11_{0.01}$ & $1.59_{0.01}$ & $1.53_{0.01}$ & $1.48_{0.01}$ & $1.63_{0.01}$ & $1.72_{0.01}$ & $2.01_{0.01}$ & $2.40_{0.01}$ \\
    & RSLS & $\mathbf{1.28}_{0.01}$ & $\mathbf{1.61}_{0.01}$ & $1.39_{0.01}$ & $1.81_{0.01}$ & $2.09_{0.01}$ & $1.60_{0.01}$ & $1.49_{0.01}$ & $1.44_{0.01}$ & $1.84_{0.01}$ & $1.71_{0.01}$ & $2.09_{0.01}$ & $2.42_{0.01}$ \\
    & LCV & $1.32_{0.01}$ & $1.63_{0.01}$ & $1.40_{0.01}$ & $1.85_{0.01}$ & $2.09_{0.01}$ & $1.57_{0.01}$ & $1.52_{0.01}$ & $1.46_{0.01}$ & $\mathbf{1.60}_{0.01}$ & $1.70_{0.01}$ & $\mathbf{1.99}_{0.01}$ & $\mathbf{2.39}_{0.01}$ \\
    
    \midrule
    
    \multirow{5}{*}{\centering\makecell[l]{$d=4$, \\ $n=100$ \\ [3pt] ($\times 0.1$)}}

    & MIAE & $3.31_{0.01}$ & $4.12_{0.01}$ & $3.55_{0.01}$ & $4.37_{0.01}$ & $5.10_{0.01}$ & $3.99_{0.01}$ & $4.20_{0.01}$ & $3.97_{0.01}$ & $4.12_{0.01}$ & $4.43_{0.01}$ & $5.25_{0.01}$ & $6.36_{0.01}$\\
    & EMI & $\mathbf{3.32}_{0.01}$ & $\mathbf{4.13}_{0.01}$ & $\mathbf{3.59}_{0.01}$ & $\mathbf{4.43}_{0.01}$ & $5.21_{0.01}$ & $\mathbf{4.06}_{0.01}$ & $4.53_{0.02}$ & $4.06_{0.01}$ & $4.23_{0.01}$ & $4.69_{0.01}$ & $5.27_{0.01}$ & $\mathbf{6.36}_{0.01}$ \\
    & LSCV & $3.43_{0.02}$ & $4.23_{0.01}$ & $3.68_{0.01}$ & $4.49_{0.01}$ & $5.25_{0.01}$ & $4.14_{0.01}$ & $4.26_{0.01}$ & $4.00_{0.01}$ & $4.15_{0.01}$ & $4.49_{0.01}$ & $5.30_{0.01}$ & $6.44_{0.01}$ \\
    & RSLS & $3.32_{0.01}$ & $4.16_{0.01}$ & $3.61_{0.01}$ & $4.45_{0.01}$ & $5.25_{0.01}$ & $4.08_{0.01}$ & $\mathbf{4.24}_{0.01}$ & $\mathbf{3.97}_{0.01}$ & $\mathbf{4.12}_{0.01}$ & $4.53_{0.01}$ & $5.31_{0.01}$ & $6.39_{0.01}$ \\
    & LCV & $3.37_{0.01}$ & $4.17_{0.01}$ & $3.64_{0.01}$ & $4.45_{0.01}$ & $\mathbf{5.15}_{0.01}$ & $4.09_{0.01}$ & $\mathbf{4.25}_{0.01}$ & $3.99_{0.01}$ & $4.14_{0.01}$ & $\mathbf{4.47}_{0.01}$ & $\mathbf{5.25}_{0.01}$ & $6.40_{0.01}$ \\
        
    \midrule

    \multirow{5}{*}{\centering\makecell[l]{$d=4$, \\ $n=500$ \\ [3pt] ($\times 0.1$)}}
    
    & MIAE & $2.27_{0.01}$ & $2.86_{0.01}$ & $2.43_{0.01}$ & $3.10_{0.01}$ & $3.65_{0.01}$ & $2.80_{0.01}$ & $3.02_{0.01}$ & $2.85_{0.01}$ & $3.05_{0.01}$ & $3.24_{0.01}$ & $3.90_{0.01}$ & $4.53_{0.01}$\\
    & EMI & $\mathbf{2.27}_{0.01}$ & $\mathbf{2.86}_{0.01}$ & $\mathbf{2.43}_{0.01}$ & $3.16_{0.01}$ & $3.80_{0.01}$ & $\mathbf{2.83}_{0.01}$ & $3.05_{0.01}$ & $\mathbf{2.85}_{0.01}$ & $3.29_{0.01}$ & $3.28_{0.01}$ & $3.95_{0.01}$ & $\mathbf{4.54}_{0.01}$ \\
    & LSCV & $2.31_{0.01}$ & $2.89_{0.01}$ & $2.47_{0.01}$ & $3.18_{0.01}$ & $3.81_{0.01}$ & $2.86_{0.01}$ & $3.05_{0.01}$ & $2.87_{0.01}$ & $3.07_{0.01}$ & $3.25_{0.01}$ & $3.92_{0.01}$ & $4.57_{0.01}$ \\
    & RSLS & $\mathbf{2.27}_{0.01}$ & $2.87_{0.01}$ & $\mathbf{2.43}_{0.01}$ & $3.17_{0.01}$ & $3.83_{0.01}$ & $2.83_{0.01}$ & $3.13_{0.01}$ & $2.87_{0.01}$ & $3.45_{0.01}$ & $3.35_{0.01}$ & $4.06_{0.01}$ & $4.55_{0.01}$ \\
    & LCV & $2.28_{0.01}$ & $2.87_{0.01}$ & $2.46_{0.01}$ & $\mathbf{3.13}_{0.01}$ & $\mathbf{3.66}_{0.01}$ & $2.84_{0.01}$ & $\mathbf{3.03}_{0.00}$ & $2.85_{0.01}$ & $\mathbf{3.05}_{0.00}$ & $\mathbf{3.24}_{0.01}$ & $\mathbf{3.90}_{0.01}$ & $4.56_{0.01}$ \\
    \bottomrule
    \end{tabular}
\end{sidewaystable}

\begin{sidewaystable}[p]
    \centering
    \caption{\small MIAE values for the first-derivative estimation simulations, with Monte Carlo standard errors shown as subscripts. Boldface indicates the minimum and values not significantly different from it among the data-driven selectors in each scenario. The oracle benchmark is the MIAE selector.}
    \label{tab:MIAE_r1}

    \begin{tabular}{llcccccccccccc}
    \toprule
    Setting & Selector & \multicolumn{6}{c}{Dirichlet} & \multicolumn{6}{c}{Gaussian-ilr} \\
    \cmidrule(lr){3-8} \cmidrule(lr){9-14}
    & & (1) & (2) & (3) & (4) & (5) & (6) & (7) & (8) & (9) & (10) & (11) & (12) \\
    \midrule
    
    \multirow{4}{*}{\centering\makecell[l]{$d=2$, \\ $n=100$}}
    
    & MIAE & $4.24_{0.03}$ & $7.03_{0.03}$ & $4.97_{0.03}$ & $8.63_{0.04}$ & $10.73_{0.03}$ & $6.41_{0.02}$ & $6.80_{0.02}$ & $6.72_{0.02}$ & $7.24_{0.02}$ & $8.39_{0.02}$ & $12.61_{0.03}$ & $16.99_{0.05}$\\
    & EMI & $4.31_{0.03}$ & $\mathbf{7.14}_{0.03}$ & $\mathbf{5.20}_{0.03}$ & $\mathbf{8.93}_{0.04}$ & $\mathbf{11.36}_{0.04}$ & $7.18_{0.05}$ & $\mathbf{7.27}_{0.03}$ & $8.21_{0.05}$ & $9.19_{0.01}$ & $10.60_{0.06}$ & $13.53_{0.06}$ & $\mathbf{17.76}_{0.08}$ \\
    & LSCV & $8.16_{0.29}$ & $10.60_{0.27}$ & $7.54_{0.23}$ & $12.79_{0.34}$ & $14.43_{0.30}$ & $9.60_{0.27}$ & $10.53_{0.33}$ & $9.47_{0.24}$ & $9.42_{0.21}$ & $10.99_{0.21}$ & $18.32_{0.50}$ & $20.14_{0.26}$ \\ 
    & RSLS & $\mathbf{4.29}_{0.03}$ & $7.18_{0.03}$ & $5.24_{0.03}$ & $9.07_{0.04}$ & $11.63_{0.04}$ & $\mathbf{6.85}_{0.03}$ & $7.37_{0.04}$ & $\mathbf{7.07}_{0.03}$ & $\mathbf{7.55}_{0.02}$ & $\mathbf{8.91}_{0.03}$ & $\mathbf{13.19}_{0.04}$ & $17.92_{0.07}$ \\

    \midrule

    \multirow{4}{*}{\centering\makecell[l]{$d=2$, \\ $n=500$}}
    
    & MIAE & $2.89_{0.02}$ & $4.88_{0.02}$ & $3.43_{0.01}$ & $6.01_{0.02}$ & $7.80_{0.02}$ & $4.59_{0.01}$ & $5.13_{0.01}$ & $4.97_{0.02}$ & $5.41_{0.02}$ & $6.23_{0.02}$ & $9.65_{0.02}$ & $12.24_{0.03}$\\
    & EMI & $2.93_{0.01}$ & $4.89_{0.02}$ & $\mathbf{3.49}_{0.01}$ & $\mathbf{6.07}_{0.02}$ & $\mathbf{8.11}_{0.02}$ & $\mathbf{4.71}_{0.01}$ & $\mathbf{5.31}_{0.02}$ & $\mathbf{5.13}_{0.02}$ & $6.70_{0.05}$ & $\mathbf{6.34}_{0.02}$ & $\mathbf{9.87}_{0.04}$ & $\mathbf{12.54}_{0.03}$ \\
    & LSCV & $4.65_{0.14}$ & $6.36_{0.12}$ & $4.80_{0.11}$ & $8.09_{0.16}$ & $9.20_{0.11}$ & $5.60_{0.08}$ & $7.12_{0.15}$ & $6.05_{0.10}$ & $\mathbf{6.22}_{0.07}$ & $7.48_{0.11}$ & $12.18_{0.19}$ & $13.56_{0.11}$ \\
    & RSLS & $\mathbf{2.89}_{0.01}$ & $\mathbf{4.88}_{0.02}$ & $3.52_{0.01}$ & $6.07_{0.02}$ & $8.16_{0.02}$ & $4.77_{0.02}$ & $5.32_{0.02}$ & $5.23_{0.02}$ & $6.37_{0.02}$ & $6.47_{0.02}$ & $10.21_{0.03}$ & $12.81_{0.03}$ \\
    
    \midrule
    
    \multirow{4}{*}{\centering\makecell[l]{$d=4$, \\ $n=100$}}

    & MIAE & $13.31_{0.06}$ & $21.03_{0.06}$ & $15.97_{0.05}$ & $25.72_{0.06}$ & $32.73_{0.06}$ & $20.11_{0.05}$ & $26.69_{0.04}$ & $24.88_{0.04}$ & $26.03_{0.03}$ & $28.65_{0.04}$ & $40.91_{0.05}$ & $68.12_{0.11}$\\
    & EMI & $\mathbf{13.36}_{0.05}$ & $\mathbf{21.10}_{0.06}$ & $\mathbf{16.47}_{0.06}$ & $\mathbf{26.36}_{0.06}$ & $\mathbf{35.14}_{0.08}$ & $\mathbf{20.88}_{0.06}$ & $30.26_{0.13}$ & $27.00_{0.10}$ & $28.51_{0.03}$ & $32.00_{0.10}$ & $43.57_{0.11}$ & $\mathbf{68.81}_{0.10}$ \\
    & LSCV & $15.03_{0.16}$ & $23.19_{0.17}$ & $17.86_{0.24}$ & $28.24_{0.27}$ & $36.62_{0.20}$ & $22.39_{0.19}$ & $28.12_{0.21}$ & $26.54_{0.16}$ & $27.09_{0.09}$ & $30.02_{0.11}$ & $45.99_{0.40}$ & $71.55_{0.22}$ \\
    & RSLS & $\mathbf{13.36}_{0.05}$ & $21.30_{0.06}$ & $16.50_{0.05}$ & $26.49_{0.06}$ & $35.41_{0.08}$ & $\mathbf{20.90}_{0.06}$ & $\mathbf{26.91}_{0.09}$ & $\mathbf{25.22}_{0.06}$ & $\mathbf{26.14}_{0.03}$ & $\mathbf{29.27}_{0.07}$ & $\mathbf{43.37}_{0.09}$ & $69.20_{0.09}$ \\
        
    \midrule

    \multirow{4}{*}{\centering\makecell[l]{$d=4$, \\ $n=500$}}
    
    & MIAE & $9.70_{0.03}$ & $15.57_{0.03}$ & $12.00_{0.03}$ & $20.04_{0.04}$ & $25.74_{0.04}$ & $15.37_{0.03}$ & $21.73_{0.03}$ & $20.19_{0.03}$ & $21.67_{0.02}$ & $23.53_{0.03}$ & $34.77_{0.03}$ & $54.66_{0.07}$\\
    & EMI & $9.71_{0.03}$ & $\mathbf{15.64}_{0.03}$ & $\mathbf{12.14}_{0.03}$ & $\mathbf{20.64}_{0.03}$ & $\mathbf{28.37}_{0.03}$ & $15.77_{0.03}$ & $\mathbf{21.77}_{0.03}$ & $\mathbf{20.36}_{0.03}$ & $25.71_{0.08}$ & $\mathbf{23.85}_{0.03}$ & $\mathbf{36.73}_{0.10}$ & $\mathbf{54.67}_{0.05}$ \\
    & LSCV & $10.56_{0.07}$ & $16.35_{0.07}$ & $13.11_{0.24}$ & $21.90_{0.29}$ & $28.94_{0.09}$ & $16.45_{0.07}$ & $22.92_{0.18}$ & $20.70_{0.04}$ & $\mathbf{22.10}_{0.03}$ & $24.18_{0.06}$ & $38.45_{0.18}$ & $56.89_{0.12}$ \\
    & RSLS & $\mathbf{9.70}_{0.03}$ & $15.68_{0.03}$ & $\mathbf{12.14}_{0.03}$ & $20.75_{0.03}$ & $28.58_{0.03}$ & $\mathbf{15.75}_{0.03}$ & $22.32_{0.03}$ & $20.82_{0.03}$ & $23.99_{0.03}$ & $24.51_{0.04}$ & $37.68_{0.05}$ & $54.69_{0.05}$ \\
    
    \bottomrule
    \end{tabular}
\end{sidewaystable}

\bibliographystyle{apalike}
\bibliography{ref}

\end{document}